\documentclass[twocolumn,showpacs]{revtex4-1}

\usepackage{graphicx}
\usepackage{color}

\begin{document}  

\title{Many-body effects, orbital mixing and cyclotron resonance in bilayer graphene}
\author{K. Shizuya}
\affiliation{Yukawa Institute for Theoretical Physics\\
Kyoto University,~Kyoto 606-8502,~Japan }

\begin{abstract}

In a magnetic field bilayer graphene supports, at the lowest Landau level, 
eight characteristic zero-energy levels with an extra twofold degeneracy in Landau orbitals $n=\{0,1\}$.
They, under general one-body and many-body interactions, evolve into pseudo-zero-mode (PZM) levels. 
A close look is made into the detailed structure and characteristics of such PZM levels 
and cyclotron resonance they host, 
with full account taken of spin splitting, interlayer bias, weak electron-hole asymmetry 
and Coulomb interactions.
It is pointed out that the PZM levels generally undergo orbital level mixing (in one valley or both valleys) 
as they are gradually filled with electrons and 
that an observation of interband cyclotron resonance over a finite range of filling 
provides a direct and sensitive probe for exploring many-body effects 
and orbital mixing in bilayer graphene. 
\end{abstract} 


\maketitle

\section{Introduction}

 Graphene supports as charge carriers  massless Dirac fermions 
which display unique and fascinating electronic properties.
Recently, increasing attention is directed to bilayers~\cite{NMMKF,MF} 
and fewlayers of graphene, where the physics and applications of graphene become far richer, 
with, e.g., a tunable band gap~\cite{MF,OBSHR}  in bilayer graphene.

In a magnetic field, graphene leads to a $\lq\lq$relativistic" infinite tower of 
electron and hole Landau levels, with the lowest Landau level (LLL) consisting of 
a number of zero-energy levels.
The presence of such zero-mode levels has a topological origin 
in the index of (the leading part of) the Dirac operators, or in the chiral anomaly~\cite{NS}. 
Monolayer graphene carries $2_{\rm valley}\times 2_{\rm spin}=4$ zero-mode levels.
Bilayer graphene supports an octet of such levels, with  an extra twofold degeneracy 
in Landau orbitals $n=0$ and $n=1$. 
In real samples, topological zero-energy levels evolve, by an interplay of 
general spin and valley splittings and Coulomb interactions,
into a variety of pseudo-zero-mode (PZM) levels, 
or broken-symmetry quantum Hall states, 
as discussed theoretically~\cite{BCNM,KSpzm,BCLM,CLBM,CLPBM,NL,MAF}.
Meanwhile it was noted that orbital degeneracy is also lifted by Coulomb interactions alone~\cite{KS_Ls}: 
Quantum fluctuations of the valence band split, 
like the Lamb shift in the hydrogen atom, the $n=0$ and $n=1$ levels appreciably. 
The orbital degeneracy and its lifting by the orbital Lamb shift are new features specific 
to the LLL in fewlayer graphene.
Experimentally, partial or full resolution of the eightfold degeneracy 
of the LLL in bilayer graphene has been observed 
in transport~\cite{FMY,ZCZJ,WAFM,MEM,VJB}  and capacitance~\cite{MFW,KFLH,HLZW} 
measurements, but the detailed structure of the LLL remains yet to be clarified.

Graphene supports cyclotron resonance (CR) in a variety of channels, 
both intraband and interband~\cite{AbF}.
Interband resonance is specific to Dirac electrons, 
with a number of competing resonance channels simultaneously activated 
over a certain range of the total Landau-level filling factor  $\nu$.
CR provides a useful means to look into the many-body problem in graphene.
In experiment CR has been explored 
for monolayer~\cite{JHTWS,DCNN,HCJL} and bilayer~\cite{HJTS,OFBB} graphene.  
Many-body effects on CR~\cite{IWF,BM,KS_CR,RFG,KS_CR_eh, ST,KS_CRnewGR,SL} 
were first detected rather indirectly by a comparison of 
some leading intraband and interband resonances~\cite{JHTWS}.  
Recently, Russell {\it et al.}~\cite{RZTW} have reported a direct signal 
of many-body effects on CR in high-mobility hBN-encapsulated monolayer graphene; 
see also ref.~\cite{prepH}. 
They observed significant variations of resonance energies 
over a finite range of filling factor $\nu$ under a fixed magnetic field $B$; 
the point lies in extracting genuine Coulombic effects as a function of $\nu$
while controlling the running of the velocity factor with $B$. 
The data show a profile nearly symmetric in $\nu$
and suggest a small band gap~\cite{HuntYYY} 
due to weak coupling to the hBN substrate.

No such $\nu$-dependent study of interband CR is yet available for bilayer graphene.
The purpose of this paper is to examine, in anticipation of a future experiment, 
the detailed characteristics of the LLL in bilayer graphene, 
with full account taken of spin splitting, interlayer bias, weak electron-hole asymmetry 
and Coulomb interactions, and to show how they are detected 
by a $\nu$-dependent and fixed-$B$ survey of interband channels of CR.
Particular attention is paid to electron-hole ($e$-$h$) conjugation and valley-interchange operations 
which relate the electron and hole bands within a valley or between the two valleys.
We clarify how they govern the Landau-level and CR spectra.

In Sec.~II we review some basic features of the effective theory of 
bilayer graphene in a magnetic field.
In Sec.~III we consider Coulombic corrections to Landau levels and CR.
Such corrections inevitably contain ultraviolet divergences and,  
in Sec.~IV, we carry out renormalization to extract observable many-body contributions.
In Secs.~V and~VI, we examine how the PZM level spectra evolve with filling, 
note the general presence of orbital mixing 
and discuss how the associated many-body effects are revealed by interband CR.  
Section VII is devoted to a summary and discussion.

\section{bilayer graphene}

The electrons in bilayer graphene are described 
by four-component spinor fields on the four inequivalent sites 
$(A,B)$ and $(A',B')$ in the Bernal-stacked bottom and top layers 
of honeycomb lattices of carbon atoms.
Their low-energy features are governed 
by the two Fermi points $K$ and $K'$ in the Brillouin zone. 
The intralayer coupling
$\gamma_{0} \equiv \gamma_{AB} \sim 3$\, eV
is related to the Fermi velocity 
$v = (\sqrt{3}/2)\, a_{\rm L}\gamma_{0}/\hbar \sim 10^{6}$~m/s 
($a_{\rm L}=  0.246$nm) in monolayer graphene.
Interlayer hopping via the dimer coupling
$\gamma_{1} \equiv \gamma_{A'B} \sim 0.4$\,eV 
makes the spectra quasi-parabolic~\cite{MF} 
in the low-energy branches $|\epsilon| <\gamma_{1}$.

The bilayer Hamiltonian with $\gamma_{0}$ and $\gamma_{1}$ 
leads to $e$-$h$ symmetric energy spectra.
Infrared spectroscopy, however, detected weak asymmetry~\cite{ZLBF,LHJ_asym}
due to the $(A,B)$-sublattice energy difference 
$\Delta \sim 18$meV and the nonleading interlayer coupling 
$\gamma_{4}\equiv \gamma_{AA'} \sim 0.04 \gamma_{0}$.

The effective Hamiltonian with such intra- and inter-layer couplings 
is written as~\cite{MF,NCGP}
\begin{eqnarray}
H &=&\! \int\! d^{2}{\bf x}\, \Big[ (\Psi^{K})^{\dag}\, {\cal H}_{K} \Psi^{K}
+ (\Psi^{K'})^{\dag}\, {\cal H}_{K'}\, \Psi^{K'}\Big], \nonumber\\
{\cal H}_{K} &=&\!\! \left(
\begin{array}{cccc}
{1\over{2}}u & 0 &-v_{4}\,  p^{\dag}   & v\,p^{\dag} \\
 0 & -{1\over{2}}u & v\,p & -v_{4}\,  p  \\
-v_{4}\,  p & v\,  p^{\dag} & \Delta  -{1\over{2}}u & \gamma_{1} \\
v\,p & -v_{4}\, p^{\dag}  & \gamma_{1}  & \Delta + {1\over{2}}u \\
\end{array}
\right),
\label{Hbilayer}
\end{eqnarray}
with 
$p= p_{x}+ i\, p_{y}$, $p^{\dag}= p_{x} - i\, p_{y}$ and
$v_{4} \equiv (\gamma_{4}/\gamma_{0})\, v$.
Here $\Psi^{K} = (\psi_{B'}, \psi_{A}, \psi_{B}, \psi_{A'})^{\rm t}$
denotes the electron field in valley $K$, with $A$ and $B$ 
referring to the associated sublattices; 
$u$ stands for an interlayer bias, 
which opens a tunable valley gap~\cite{OBSHR,MF}. 
We ignore the effect of trigonal warping 
$\propto \gamma_{3}\equiv \gamma_{AB'} \sim 0.1$\,eV 
which, in a strong magnetic field, causes 
only a negligibly small level shift.
${\cal H}_{K}$ is diagonal in the (suppressed) electron spin.

As a typical set of band parameters we adopt those by recent theoretical calculations~\cite{JM,fn}, 
\begin{eqnarray}
v &\approx&0.845 \times 10^{6}\, {\rm m/s},\  
\gamma_{1} \approx 361\, {\rm meV}, \nonumber\\
v_{4}/v &\equiv& \gamma_{4}/\gamma_{0}\approx 0.053,\ 
\triangle \approx  15\, {\rm meV},
\label{expparameters}
\end{eqnarray}
which are also used in an experimental analysis~\cite{HLZW}.

Let us note that ${\cal H}_{K}$ is unitarily equivalent to 
$-{\cal H}_{K}$ with the signs of $(u, \Delta, v_{4})$ 
reversed, 
\begin{equation}
S^{\dag}{\cal H}_{K}\, S = - {\cal H}_{K}|_{-u, -\Delta, -v_{4}},
\end{equation}
with $S= {\rm diag}(-1,1,-1,1)$;
${\cal O}|_{-u, -\Delta, -v_{4}}$ signifies setting 
$(u, \Delta, v_{4}) \rightarrow (-u, -\Delta, -v_{4})$ in ${\cal O}$.
The electron and hole bands are therefore intimately related within a valley.

The Hamiltonian ${\cal H}_{K'}$ in another valley is given by ${\cal H}_{K}$ 
with $(v, v_{4}, u) \rightarrow (-v,-v_{4}, -u)$, 
and acts on a spinor of the form 
$\Psi^{K'} = (\psi_{A},\psi_{B'}, \psi_{A'}, \psi_{B})^{\rm t}$.
Actually, ${\cal H}_{K'}$ is unitarily equivalent to 
${\cal H}_{K}$ with the sign of $u$ reversed,  
\begin{equation}
{\cal H}_{K'} ={\cal U}^{\dag}{\cal H}_{K}|_{-u}\,{\cal U},
\label{valleyConjugation}
\end{equation}
with ${\cal U}= {\rm diag}(1,1,-1,-1)$.
In what follows we adopt ${\cal H}_{K}|_{-u}$ for ${\cal H}_{K'}$
and simply pass to valley $K'$ by reversing the sign of bias $u$ in valley-$K$ expressions.
We also suppose, without loss of generality, that $u \ge 0$ in valley $K$; 
thus $u \le 0$ refers to valley $K'$.

The unitary equivalence
\begin{equation}
{\cal H}_{K'}\stackrel{\cal U}{\sim}  {\cal H}_{K}|_{-u}
 \stackrel{S}{\sim} - {\cal H}_{K}|_{-\Delta, -v_{4}}
 \label{unitary_equiv}
 \end{equation}
implies that the $e$-$h$ conjugation symmetry
is kept exact only for $\Delta = v_{4} =0$ 
although, if $u\not=0$, it is apparently broken within each valley.
This symmetry analysis also holds in the presence of Coulomb interactions.

Let us place bilayer graphene in a strong uniform magnetic field 
$B_{z} = B>0$ normal to the sample plane;
we set, in ${\cal H}_{K}$, 
$p\rightarrow \Pi = p + eA$
with $A= A_{x}+ iA_{y}= -B y$, and rescale
$\Pi = -(\sqrt{2}/\ell)\, Z^{\dag}$ so that 
$[Z, Z^{\dag}]=1$, where $\ell=1/\sqrt{eB}$ denotes  the magnetic length.

The eigenmodes of ${\cal H}_{K}$ are labeled by integers $N\equiv |n| =(0,1,2,\cdots)$ 
and plane waves with momentum $p_{x}$, and  have the structure 
\begin{equation}
\Psi^{n} = \Big(|N\!-\!2\rangle\, b'_{n} ,|N\rangle\, c_{n},
|N\!-\!1\rangle\, b_{n}, |N\!-\!1\rangle\, c'_{n} \Big)^{\rm t}, 
\end{equation} 
where only the orbital modes are shown 
using the standard harmonic-oscillator basis $\{ |N\rangle \}$, 
with $|N\rangle =0$ for $N<0$. 
The coefficients ${\bf b}_{n}=(b'_{n}, c_{n}, b_{n}, c'_{n})^{\rm t}$
for each $N$ are given by the normalized eigenvectors 
of the reduced Hamiltonian
\begin{equation}
\hat{\cal H}_{N} = \omega_{c} \left(
\begin{array}{cccc}
\mu & 0  &  r\,  \sqrt{N\!-\!1} &  -\sqrt{N\!-\!1} \\
0 & - \mu & -\sqrt{N} & r\, \sqrt{N}  \\
r\,  \sqrt{N\!-\!1}  & -\sqrt{N}  & d - \mu & g \\
 -\sqrt{N\!-\!1} & r\, \sqrt{N}  &g &d+ \mu \\
\end{array}
\right),
\label{reducedH}
\end{equation} 
where 
\begin{equation}
\omega_{c}\equiv \sqrt{2}\, v/\ell 
\approx 36.3 \times v[10^{6}{\rm m/s}]\, \sqrt{B[{\rm T}]}\ {\rm meV},
\end{equation} 
is the cyclotron energy 
for monolayer graphene; $g\equiv \gamma_{1}/\omega_{c}$,
$\mu \equiv {1\over{2}}\, u/\omega_{c}$, $d\equiv \Delta/\omega_{c}$ and 
$r\equiv \gamma_{4}/\gamma_{0}$.  
Numerically, for the set of parameters in Eq.~(\ref{expparameters}) at $B=20$\,T,
\begin{equation}
\omega_{c}\approx 137\,{\rm meV},\
(g, d,r) \approx (2.63, 0.109, 0.053), 
\label{Param}
\end{equation} 
and $\mu\approx 0.073$ for $u=20$meV.
In what follows, we employ this set~(\ref{Param}) of parameters for numerical estimates at $B=20$\,T.

The unitary equivalence in Eq.~(\ref{unitary_equiv})  reveals 
how the spectra $\epsilon_{n}$ and the associated eigenmodes ${\bf b}_{n}$
change via $e$-$h$ conjugation.
Within each valley they read
\begin{eqnarray}
&& \epsilon_{-n} = -\epsilon_{n}|_{-u, -\Delta, -v_{4}} ,
\nonumber\\
&& (b'_{-n}, c_{-n}, b_{-n}, c'_{-n}) 
= (- b'_{n}, c_{n}, -b_{n}, c'_{n})|_{-u, -\Delta, -v_{4}},\ \ \ \
\label{EC_each}
\end{eqnarray}
while, between the two valleys, they are related as  
\begin{eqnarray}
&& ( \epsilon_{n}; b'_{n}, c_{n}, b_{n}, c'_{n})|^{K'} 
=( \epsilon_{n}; b'_{n}, c_{n}, b_{n}, c'_{n})|^{K}_{-u} 
\nonumber\\
&&= (-\epsilon_{-n};  b'_{-n}, c_{-n}, -b_{-n}, c'_{-n})|^{K}_{-\Delta, -v_{4}}.
\label{EC_two}
\end{eqnarray}

Of our particular concern are the $n=0$ and $n=1$ modes.
For $n=0$, $\hat{\cal H}_{N=0}$ has an obvious eigenvalue 
and eigenvector
\begin{equation}
\epsilon_{0}= - u/2 = -\omega_{c}\mu, \ \ {\bf b}_{0} = (0,1,0,0)^{\rm t}.
\end{equation}
For $N=1$, $\hat{\cal H}_{N}$ has three solutions
$(\epsilon_{-1'}, \epsilon_{1},\epsilon_{1'})$,
with $\epsilon_{\pm1'} \sim \pm \gamma_{1}$ belonging to the higher branches.

The $n=1$ mode (with $\epsilon_{1}$) and the $n=0$ mode 
have zero energy for $(u, \Delta, r) =0$ and deviate from zero energy as $(u, \Delta, r)$ develop.
These pseudo-zero modes (PZM) form the LLL in bilayer graphene.
The eigenenergy and eigenvector of the $n=1$ mode are written as 
\begin{eqnarray}
\epsilon_{1} &=& \omega_{c} (-\mu +\kappa),
\nonumber\\
{\bf b}_{1}&=& 
c_{1} \Big(0,1, - (  g\kappa - r)/\tilde{g} , (1+ \kappa d - d^2)/\tilde{g} \Big),\ \ \ 
\nonumber\\
\tilde{g} &=& g+ (d- \kappa)\, r.
\label{PZMlevels}
\end{eqnarray}
In what follows, we denote valley and $e$-$h$ symmetry breaking collectively as 
$X =(u, \Delta, v_{4})$ or $X=(\mu, d, r)$.
 As seen from the secular equation, the correction 
$\kappa \equiv \kappa(g; \mu,d,r)$ in $\epsilon_{1}$ is odd in $X$, 
and hence the normalization factor $c_{1}$ is even in $X$, 
\begin{eqnarray}
\kappa &\approx& {2\mu+ d + 2r\, g \over{ g^2 +1}} + O(X^3),
\nonumber\\
c_{1} &\approx &  1/\sqrt{1 + (1/g^2)} + O(X^2).
\label{energyN1}
\end{eqnarray}
Numerically, at $B=20$\,T,  $\kappa \approx 0.049+ 0.25\, \mu$ 
while $c_{1} \approx 0.934 -0.007\,  \mu - 0.02\, \mu^2$ scarcely depends on $\mu$
[for small $\mu \sim O(0.1)$].

Interlayer bias $u\propto \mu$, if nonzero, breaks valley symmetry 
and shifts the PZM levels $n=\{0,1\}$ 
oppositely ($\propto \mp u/2$) in the two valleys. 
Accordingly, $(\epsilon_{0}, \epsilon_{1})$ are nearly degenerate in each valley 
and are ordered so that 
$\epsilon_{0}^{K} \lesssim \epsilon_{1}^{K} < 0 
< \epsilon_{1}^{K'} \lesssim \epsilon_{0}^{K'}$ for $\mu >0$.

For $N\ge 2$, $\hat{\cal H}_{N}$ has rank 4 and 
we denote the four branches of  Landau-level spectra as 
$\epsilon_{-n'} < \epsilon_{-n}<0<\epsilon_{n} < \epsilon_{n'}$ 
(with $|\epsilon_{\pm n'}| \gtrsim \gamma_{1}$)
so that the index $(n, n')$ reflects the sign of $\epsilon_{n}$. 
Let us denote $\epsilon_{n}$ in units of $\omega_{c}$, 
\begin{equation}
\epsilon_{n} = \omega_{c}\, e_{n}\ {\rm and}\ \epsilon_{n'} = \omega_{c}\, e_{n'}.
\end{equation}
For  $X=(\mu, d, r)\rightarrow 0$, the (dimensionless) spectra 
$e_{n} \equiv e_{n}(g; \mu,d,r)$ of the lower branches take the form 
\begin{equation}
e_{n}^{(0)}
=s_{n}\sqrt{  {2|n| (|n|-1)\over{g^2 + a_{n} + \sqrt{D_{n}}}} } \sim O(1/g),
\label{eta_nzero}
\end{equation}
where $e_{n}^{(0)} \equiv e_{n}|_{X=0}$,  
$D_{n}= g^4 + 2 a_{n}g^2 +1$ and $a_{n}= 2\, |n|-1$;
$s_{n} \equiv {\rm sgn}[n] = \pm 1$ is the sign function.
The full spectra, to first order in breaking $X$, are written as 
\begin{equation}
e_{n} \approx  e_{n}^{(0)} 
 +{1\over{2}} (1-{g^2 \over{\sqrt{D_{n}}}})\,  d
  +{ \mu + a_{n}\, r\, g \over{\sqrt{D_{n}}}} .
\end{equation}
Actually, $e$-$h$ breaking 
$(d,r)$, listed in Eq.~(\ref{Param}), has a sizable effect 
and shifts all the spectra, except for $\epsilon_{0}$, upwards appreciably 
[e.g., roughly by 10 \% for $n=(2, 3)$], as seen from Fig.~1, 
which, for later convenience, is presented in Sec.~V.
Unlike $e_{0}\sim e_{1} \sim O(\mu)$,  
$e_{n}$ depend on bias $u$ only weakly $\sim O(\mu/g^2)$.
The spectra $\epsilon_{n'}$ 
of the higher branches $n' =\pm1', \pm 2', \dots$
show similar but partially different dependence on breaking $X$.

Each $|n| \ge 2$ is associated with a pair of electron and hole modes 
$\pm n$  (and $\pm n'$). In contrast, the pseudo-zero modes $n=0$ and $n=1$ 
stand alone (per spin and valley) and are, in this sense, $e$-$h$ selfconjugate, 
with $\pm n \rightarrow n$ in Eqs.~(\ref{EC_each}) and (\ref{EC_two}). 
Thus $\epsilon_{0} = -\epsilon_{0}|_{-X}= -\omega_{c}\mu$ 
and $\epsilon_{1} = -\epsilon_{1}|_{-X}$; that is,  
$\epsilon_{0} = \epsilon_{1}= 0$ for $X \rightarrow 0$ 
and $\epsilon_{1}$ consists of odd powers of breaking $X$.

The Landau-level structure is made explicit by passing to
the $|n,y_{0}\rangle$ basis  (with $y_{0}\equiv \ell^{2}p_{x}$) 
via the expansion
$\Psi^{a}_{\alpha} ({\bf x}) = \sum_{n, y_{0}} \langle {\bf x}| n, y_{0}\rangle\, 
\psi^{n;a}_{\alpha}(y_{0})$, 
where $n$ refers to the Landau level index, 
$\alpha \in (\uparrow, \downarrow)$ 
to the spin and $a \in (K,K')$ to the valley.
It is tacitly understood that the sum is taken 
over the higher branches $n' = (\pm 1',\pm 2', \dots)$ as well.
The one-body Hamiltonian is written as
\begin{equation}
H = \int dy_{0}\sum_{m, n}\sum_{a,\alpha} 
\psi_{\alpha}^{m;a \dag}(y_{0})\, \epsilon_{n}^{a;\alpha}\, \psi_{\alpha}^{n;a}(y_{0}).
\end{equation}
The charge density 
$\rho_{-{\bf p}} =\int d^{2}{\bf x}\,  e^{i {\bf p\cdot x}}\,\rho$ 
with $\rho = (\Psi^{K})^{\dag}\Psi^{K} +  (\Psi^{K'})^{\dag}\Psi^{K'}$ 
is thereby written as~\cite{KS_CR}
\begin{eqnarray}
\rho_{-{\bf p}} &=& \gamma_{\bf p}\sum_{m, n =-\infty}^{\infty}
\sum_{a,\alpha} g^{m n;a}_{\bf p}\, 
R^{m n;aa}_{\alpha\alpha;\bf -p}, \nonumber\\
R^{m n;ab}_{\alpha\beta;{\bf -p}}&\equiv& \int dy_{0}\,
{\psi^{m;a}_{\alpha}}^{\dag}(y_{0})\, e^{i{\bf p\cdot r}}\,
\psi^{n;b}_{\beta} (y_{0}),
\label{chargeoperator}
\end{eqnarray}
where $\gamma_{\bf p} =  e^{- \ell^{2} {\bf p}^{2}/4}$; 
${\bf r} = (i\ell^{2}\partial/\partial y_{0}, y_{0})$
stands for the center coordinate with uncertainty 
$[r_{x}, r_{y}] =i\ell^{2}$.
The charge operators $R^{m n;aa}_{\alpha\alpha;\bf p}$ obey 
the $W_{\infty}$ algebra~\cite{GMP}.

The coefficient matrix $ g^{m n;a}_{\bf p} \equiv g^{mn}_{\bf p}|^{a}$ 
in valley $a\in (K,K')$ is constructed 
from the eigenvectors ${\bf b}_{n}|^{a}$,
\begin{eqnarray}
g^{mn}_{\bf p} &=& c_{m}\, c_{n}\, f_{\bf p}^{|m|,|n|}
+ b'_{m}\, b'_{n}\, f_{\bf p}^{|m|-2,|n|-2} \nonumber\\
&&+ (b_{m}\, b_{n}+ c'_{m}\, c'_{n})\, f_{\bf p}^{|m|-1,|n|-1},
\label{gkn}
\end{eqnarray}
where
\begin{equation}
f^{m n}_{\bf p} 
= \sqrt{{n!\over{m!}}}\,
\Big({- \ell p^{\dag}\over{\sqrt{2}}}\Big)^{m-n}\, L^{(m-n)}_{n}
\Big ({1\over{2}} \ell^{2}{\bf p}^{2}\Big)
\label{fknp}
\end{equation}
for $m \ge n\ge0$, and $f^{n m}_{\bf p} = (f^{m n}_{\bf -p})^{\dag}$;
$f^{m n}_{\bf p}=0$ for $m<0$ or $n<0$;
$p=p_{x}\! +i\, p_{y}$.
In view of Eqs.~(\ref{EC_each}) and ~(\ref{EC_two}), 
$g^{mn}_{\bf p}$ have the following property under $e$-$h$
conjugation,
\begin{eqnarray}
&&g^{-m,-n;a}_{\bf p} = g^{m,n;a}_{\bf p}|_{-\mu,-d,-r},
\nonumber\\
&&g^{mn}_{\bf p}|^{K'} = g^{m,n}_{\bf p}|^{K}_{-\mu} = g^{-m,-n}_{\bf p}|^{K}_{-d,-r}.
\label{gmn_eh}
\end{eqnarray}

For $n, m \in (0,1)$, $g^{m,n;a}_{\bf p} = g^{m,n;a}_{\bf p}|_{-X}$ 
are even functions of breaking $X$.
They actually take simple form   
\begin{eqnarray}
g^{00}_{\bf p} &=& 1, \ \
g^{11}_{\bf p} =1-c_{1}^{2}\, \textstyle{1\over{2}}\ell^{2}{\bf p}^{2}, 
\nonumber\\
g^{01}_{\bf p} &=& c_{1} \ell\, p/\sqrt{2},\ \ 
g^{10}_{\bf p} = -c_{1} \ell\, p^{\dag}/\sqrt{2},
\label{gmnPZM}
\end{eqnarray}
in each valley, with $c_{1}|^{K'} = c_{1}|^{K}_{-\mu}$.

The Coulomb interaction 
$V= {1\over{2}} \sum_{\bf p} v_{\bf p}\,  {:\! \rho_{\bf -p}\, \rho_{\bf p}\!:}$ 
is written as
\begin{equation}
V = {1\over{2}} \sum_{\bf p}
v_{\bf p}\,\gamma_{\bf p}^{2}\,  
g^{jk; a}_{\bf p}\, g^{m n;b}_{\bf -p}
:\! R^{j k;aa}_{\alpha\alpha;{\bf -p}}\, 
R^{m n;bb}_{\beta\beta;{\bf p}}\! : , 
\label{VCoul}
\end{equation}
with the potential $v_{\bf p}= 2\pi \alpha_{e}/(\epsilon_{\rm b} |{\bf p}|)$,
$\alpha_{e} \equiv e^{2}/(4 \pi \epsilon_{0})$ and 
the substrate dielectric constant $\epsilon_{\rm b}$;
$\sum_{\bf p} \equiv \int d^{2}{\bf p}/(2\pi)^{2}$;  $:\ :$ denotes normal ordering. 
For simplicity, we ignore a small effect of interlayer separation. 
Here and from now on, we suppress
summations over levels $n$, valleys $a$ and spins $\alpha$, 
with the convention that the sum is taken over repeated indices.
The one-body Hamiltonian $H$  is thereby written as
\begin{equation}
H = \epsilon_{n}^{a;\alpha}\, R^{nn; aa}_{\alpha\alpha;{\bf p= 0}} 
- \mu_{\rm Z}\, (T_{3})_{\alpha\beta} R^{nn;aa}_{\alpha\beta;{\bf p=0}}.
\label{Hzero}
\end{equation}
Here the Zeeman term $\mu_{\rm Z} \equiv g^{*}\mu_{\rm B}B$ is introduced 
via the spin matrix $T_{3} = \sigma_{3}/2$.

\section{Coulombic corrections}

In this section we study Coulombic contributions to the Landau-level 
and associated CR spectra.
In graphene, unlike conventional quantum Hall systems, 
the electrons and holes are always subject to quantum fluctuations 
of the infinitely-deep filled valence band (or the Dirac sea), 
which, being strong, lead to ultraviolet divergences.
One first has to handle them properly.

The Coulomb direct interaction leads to a divergent self-energy 
$\propto v_{\bf p \rightarrow 0}$, which, as usual, is removed 
when a neutralizing background is taken into account.
The exchange interaction, on the other hand, gives rise
to corrections to level spectra $\epsilon_{n}^{a; \alpha}$ of the form 
\begin{equation}
\Delta \epsilon_{n}^{a; \alpha}
= -\sum_{\bf p}v_{\bf p}\gamma_{\bf p}^{2}\, 
\sum_{m} \nu_{m}^{a; \alpha}\, |g^{n m;a}_{\bf p}|^2.
\label{self_En}
\end{equation}
Here $0\le \nu_{n}^{a; \alpha} \le 1$ stands for the filling fraction 
of the $(n, a, \alpha)$ level.
The exchange interaction preserves the spin and valley 
$(\alpha,a)$. Accordingly, from now on, we suppress them
 and mainly refer to valley $K$.  

The self-energies $\Delta \epsilon_{n}$ 
involve a sum over infinitely many filled levels $m$ in the valence band.  
Their structure is clarified if one notes the completeness relation
\begin{equation}
\sum_{k=-\infty}^{\infty}|g^{n k}_{\bf p}|^2 
=1/ \gamma_{\bf p}^{2}
= e^{{1\over{2}}\ell^2 {\bf p}^2}.
\label{completeness-rel}
\end{equation}
Actually, this follows from the fact~\cite{KS_LWGS} that  
$G^{n k}_{\bf p} \equiv \gamma_{\bf p}\, g^{n k}_{\bf p}$ 
are ($W_{\infty}$) unitary matrices that obey the composition law
$G_{\bf p}\, G_{\bf q} = e^{i{1\over{2}}\ell^2{\bf p}\times {\bf q}}\, G_{\bf p+q}$
with ${\bf p}\times {\bf q}= p_{x}q_{y}-p_{y}q_{x}$.
The half-infinite sum in $\Delta \epsilon_{n}$ is then cast in the form
\begin{eqnarray}
\gamma_{\bf p}^{2}\! \sum_{k\le -2}|g^{n k}_{\bf p}|^2 \!
&=&\! {\textstyle {1\over{2}}} -  {\textstyle {1\over{2}}}\, F_{n}(z) 
- {\textstyle {1\over{2}}} \gamma_{\bf p}^{2}\, (|g^{n 0}_{\bf p}|^2 +|g^{n 1}_{\bf p}|^2), 
\nonumber\\
F_{n}(z) &\equiv&
\gamma_{\bf p}^{2}\sum_{k=2}^{\infty}\{ |g^{nk}_{\bf p}|^2 -  |g^{n,-k}_{\bf p}|^2 \},
\end{eqnarray}
where $z= {1\over{2}}\ell^2 {\bf p}^{2}$.
Here again it is tacitly understood that the sum is taken over $k' \le -1$ or $k' \ge 1$ as well.

The self-energies $\Delta \epsilon_{n}$ are thereby rewritten as 
\begin{equation}
\Delta \epsilon_{n}  
= \sum_{\bf p}v_{\bf p} \Big[\!
- {\textstyle{1\over{2}}}+ {\textstyle{1\over{2}}}\, F_{n}(z)
-\sum_{k} \nu [k]\,  \gamma_{\bf p}^{2}  |g^{nk}_{\bf p}|^2\Big].
\end{equation}
Here the last term with the $\lq\lq$electron-hole" filling factor,
\begin{eqnarray}
\nu[k] &=& \nu_{k}\, \theta_{(k \ge 2)} -  (1-\nu_{k})\theta_{(k\le -2)}
 \nonumber\\
 &&+ (\nu_{1}- {\textstyle{1\over{2}} })\, \delta_{k,1}
 + (\nu_{0}- {\textstyle{1\over{2}} })\, \delta_{k,0},
\label{nu_k_ehsym}
\end{eqnarray}
 where 
$\theta_{(k\ge 2)} =1$ for $k\ge 2$ and $\theta_{(k\ge 2)} =0$ otherwise, etc.,
refers to a finite number of 
filled electron or hole levels around the PZM sector $n=\{0,1\}$.

Now $F_{n}(z) \equiv F_{n}(z; g, \mu,d,r)$ involve a sum over infinitely many levels.
$F_{n}(z) \propto 1/\sqrt{z}$ for $z \rightarrow \infty$ and 
yield ultraviolet divergences upon integration over ${\bf p}$ with $v_{\bf p}$. 
In view of Eq.~(\ref{gmn_eh}), $F_{\pm n}(z)$ are related in each valley or between the valleys,
\begin{eqnarray}
F_{-n}(z) &=& - F_{n}(z)|_{-\mu, -d, -r},
\nonumber\\
F_{n}(z)|^{K'} &=& F_{n}(z)|^{K}_{-\mu} = - F_{-n}(z)|^{K}_{-d, -r}.
\label{eh_Fn}
\end{eqnarray}
For $n=(0,1)$, 
$F_{n}(z) = - F_{n}(z)|_{-X}$; 
$F_{0}(z)$ and $F_{1}(z)$ 
are odd in breaking $X=(u,d,r)$. 
The PZM sector thus has no divergence for $X=0$, 
$(F_{0}(z), F_{1}(z))|_{X\rightarrow0} \rightarrow 0$.

Let us eliminate from $\Delta \epsilon_{n}$ a constant, 
$\sum_{\bf p}v_{\bf p}\, [-{1\over{2}} + {1\over{2}} F_{0}(z)|_{\mu=0}]$  common to all levels, 
which is safely done by adjusting zero of energy.
Via such regularization, self-energies $\Delta \epsilon_{n}$ are cast in the form 
\begin{equation}
\Delta \epsilon_{n}  
\stackrel{\rm reg}{=} {\textstyle{1\over{2}}} \sum_{\bf p}v_{\bf p}\, {\cal F}_{n}(z)
-\sum_{k} \nu [k]  \sum_{\bf p}v_{\bf p}\, \gamma_{\bf p}^{2}  |g^{nk}_{\bf p}|^2.
\label{DEn_reg}
\end{equation}
Note that  
\begin{equation}
{\cal F}_{n}(z) \equiv F_{n}(z) -  F_{0}(z)|_{\mu=0}.
\end{equation}
inherit the conjugation property of $F_{n}(z)$ in Eq.~(\ref{eh_Fn}).

A key property of the electron-hole filling factor $\nu[k]$ defined in Eq.~(\ref{nu_k_ehsym})
is that it is odd under $e$-$h$ conjugation:
$\nu[k]$ changes sign upon interchanging  
the electron and hole levels $(k \rightarrow -k)$ 
by replacing, in $\nu[-k]$, $\nu_{-k} \rightarrow 1- \nu_{k}$. 
Noting this feature and rewriting Eq.~(\ref{DEn_reg}) then allows us to relate, 
as done earlier~\cite{KS_CRnewGR} for monolayer graphene, 
$\Delta \epsilon_{\pm n}$ 
in a valley or between the valleys. 
In terms of the full spectra to $O(V)$,
\begin{equation}
\hat{\epsilon}_{n} = \epsilon_{n} + \Delta  \epsilon_{n},
\end{equation}
the result is 
\begin{equation}
\hat{\epsilon}_{n}^{K}|^{N_{\rm f}=m} =
-\hat{\epsilon}_{-n}^{K}|^{N_{\rm f}=2-m}_{-\mu,-d,-r}
= -\hat{\epsilon}_{-n}^{K'}|^{N_{\rm f}=2-m}_{-d,-r}.
\label{En-ehconj}
\end{equation}
Here $N_{\rm f}$ specifies filling of the relevant valley; note that  
$\Delta \epsilon_{n}$ depend on filling. 
We assign $N_{\rm f}=0$ to the empty PZM sector (of a given spin and valley) 
with levels below it ($n\le -2$) all filled,
or with the $\lq\lq$uppermost filled level" $n_{\rm f}=-2$; 
accordingly, $N_{\rm f} =(-1,0; 3,4)$, e.g., refer to $n_{\rm f}=(-3,-2;2,3)$.
$N_{\rm f} =1$ refers to the PZM sector with one filled level and 
$N_{\rm f} =2$ to the filled sector. 
For definiteness, in what follows, we focus on cases of integer filling, 
in which a distinct band gap is present, 
and take $N_{\rm f}$ in Eq.~(\ref{En-ehconj}) to be integers.
Let us now imagine, e.g., valley $K$ with $N_{\rm f} =m$. 
Interchanging electrons and holes (according to $\nu_{-k} \rightarrow 1- \nu_{k}$) 
yields a configuration with $N_{\rm f} =2-m$, and Eq.~(\ref{En-ehconj}) implies 
that the spectra of  such $e$-$h$ conjugate configurations are intimately related. 
A typical $e$-$h$ conjugate pair are the empty PZM sector ($N_{\rm f} =0$) 
and the filled one ($N_{\rm f} =2$).

Some care is needed in handling the $N_{\rm f}=1$ configuration,
which is not uniquely fixed, e.g., $N_{\rm f}=1$ with $n_{\rm f}=0$ or $n_{\rm f}=1$ 
or with any linear combination of $\{0,1\}$.
Note also that $e$-$h$ conjugation reverses the level layout of the PZM sector,
e.g., $(0,1) \rightarrow (1,0)$.
As a result, if, e.g., $\hat{\epsilon}_{0}< \hat{\epsilon}_{1}$ in a valley,  
Eq.~(\ref{En-ehconj}) reads 
\begin{equation}
(\hat{\epsilon}_{0}, \hat{\epsilon}_{1})|^{N_{\rm f}=1, n_{\rm f}=0} =
(-\hat{\epsilon}_{0}, -\hat{\epsilon}_{1})|^{N_{\rm f}=1, n_{\rm f}=1}_{-X},
\label{Nfone_NoRotation}
\end{equation}
which means that the filled $n=0$ level turns into the filled $n=1$ level 
in the conjugate configuration. 
In Sec.~V, we study the PZM sector over a continuous range $0 \le N_{\rm f} \le 2$ 
and encounter an interesting case of mixed  $\{0,1\}$ levels at $N_{\rm f}=1$.

Let us next study CR, namely, optical interlevel transitions at zero momentum transfer,
with the selection rule~\cite{AbF} $\Delta |n|=\pm 1$ for graphene, 
 i.e., 
(i) intraband channels $\{n\leftarrow n-1, -(n-1) \leftarrow -n\}$
and 
(ii) interband channels $T_{n}\equiv \{n\leftarrow -(n-1),\  n-1 \leftarrow -n\}$ for $n=1,2, \cdots$;
$T_{1} = \{ 1\! \leftrightarrow\! 0\}$,
$T_{2} = \{ 2 \leftarrow 1,\  1 \leftarrow -2\}$, $T_{3} = \{ 3 \leftarrow -2,\  2 \leftarrow -3\}$, etc.
Interband CR is specific to Dirac electrons.
Consider now CR from level $j$ to level $n$ for each (valley, spin)=$(a,\alpha)$ channel
and denote the associated excitation energy as
$\epsilon_{\rm exc}^{n\leftarrow j} = \epsilon_{n}- \epsilon_{j} +\Delta \epsilon^{n,j}$;
CR preserves the valley and spin so that $(a,\alpha)$ will be suppressed  from now on.
In the mean-field treatment~\cite{KS_CR,KH,MOG}, the corrections 
\begin{eqnarray}
\Delta \epsilon^{n,j} &=& \Delta \epsilon_{n} -\Delta \epsilon_{j} 
-  (\nu_{j}-  \nu_{n} )\,{\cal A}_{n,j} ,
\nonumber\\
{\cal A}_{n,j} 
&=& \sum_{\bf p} v_{\bf p}\gamma_{\bf p}^{2}\, g^{nn}_{\bf -p}\, g^{jj}_{\bf p},
\label{Eexc_SMA} 
\end{eqnarray}
consist of the self-energy difference 
and the Coulombic attraction ${\cal A}_{n,j}$ between the excited electron and created hole.
The full CR spectra thus differ from the dressed level gaps by attraction energies ${\cal A}_{n,j}$,
\begin{equation}
\epsilon_{\rm exc}^{n\leftarrow j} 
= \hat{\epsilon}_{n}- \hat{\epsilon}_{j} -  (\nu_{j}-  \nu_{n} )\,{\cal A}_{n,j}.
\label{fullEnj}
\end{equation}
It will be clear now that, via $e$-$h$ conjugation, 
$\epsilon_{\rm exc}^{n \leftarrow -j}$ turns into 
$\epsilon_{\rm exc}^{j \leftarrow -n}$.
In particular, the interband CR channels 
$T_{n}\equiv \{n\leftarrow -(n-1),\ n-1 \leftarrow -n\}$
are intimately related, 
\begin{eqnarray}
\epsilon_{\rm exc}^{n \leftarrow -(n-1)}|^{K;N_{\rm f}=m} \!
&=&\epsilon_{\rm exc}^{n-1 \leftarrow -n}|^{K;N_{\rm f}=2-m}_{-\mu, -d,-r},
\nonumber\\
&=& \epsilon_{\rm exc}^{n-1 \leftarrow -n}|^{K';N_{\rm f}=2-m}_{-d,-r};
\label{Eexc-ehconj}
\end{eqnarray}
we suppose $N_{\rm f} \not=1$ here. 
Such an  $\lq\lq$$e$-$h$ conjugate" character of $T_{n}$ 
was noticed earlier~\cite{KS_CRnewGR} for monolayer graphene, 
for which $e$-$h$ conjugation is an exact symmetry.  
Here for bilayer graphene  $e$-$h$ conjugation, though not a symmetry, 
is an operation that tells us how the CR spectra deviate from 
the symmetric ones by breaking $(\mu,d,r)$.
Experimentally the conjugate channels of a given set $T_{n}$ 
are observable as competing  signals over a finite range of filling factor $\nu$ 
and are indistinguishable unless polarized light is used.

\section{renormalization}

The self-energies $\Delta \epsilon_{n}$ are afflicted with ultraviolet divergences. 
In this section we consider how to extract physically observable information 
out of them.
To this end, one first defines renormalized parameters by setting 
$v = Z_{v} v^{\rm ren} = v^{\rm ren} + \delta v$, 
$\gamma_{1}= \gamma_{1}^{\rm ren} + \delta \gamma_{1}$, 
$\Delta= \Delta^{\rm ren} + \delta \Delta$, etc., with counterterms 
$\delta v= (Z_{v}-1)\,  v^{\rm ren} \sim  O(\alpha_{e})$, $\delta \gamma_{1}$, etc.
The one-body Hamiltonian $H = H^{\rm ren} + \delta_{\rm ct} H$ is then divided into 
the portion $H^{\rm ren}$ involving only the renormalized quantities and 
the counterterms  $\delta_{\rm ct} H$.
One starts with $H^{\rm ren}$, calculates the Coulombic corrections 
and encounters divergences. 
If one could remove them by adjusting $\delta_{\rm ct} H$,
renormalization is done properly.

Fortunately, the renormalization procedure 
to $O(V) \sim O(\alpha_{e})$ 
for bilayer graphene in a magnetic field $B$ has been formulated earlier~\cite{KS_CR_eh}.
The key point is that, since $B$ simply acts 
as a long-wavelength cutoff $\sim \ell$, 
the short-distance structure of the present theory 
is known from its free-space ($B=0$) version:
(i) The divergence associated with velocity $v$ is the same 
as that for monolayer graphene, 
\begin{equation}
\delta v= -(\alpha_{e}/8\epsilon_{b}) \log (\ell\Lambda)^2,
\end{equation} 
with a  momentum cutoff $\Lambda$.
 (ii)~$v_{4}$ and $u$ remain finite,
\begin{equation}
\delta v_{4}= \delta u = 0.
\end{equation} 
We thus take
$r = v_{4}/v^{\rm ren}$ and  $\mu = (u/2)/\omega_{c}^{\rm ren}$ 
 finite; $\omega_{c}^{\rm ren} \equiv \sqrt{2}v^{\rm ren}/\ell$.
(iii)  $\gamma_{1}$ and $\Delta$, though mixed under renormalization, 
are also governed by $\delta v$,
\begin{eqnarray}
\delta \gamma_{1}&=& (\gamma_{1}^{\rm ren} + r\, \Delta^{\rm ren})\, 
h(r)\, \delta v/v^{\rm ren},  \nonumber\\
\delta \Delta &=& 2\, (\Delta^{\rm ren} + r\, \gamma_{1}^{\rm ren})\,
h(r)\, \delta v/v^{\rm ren},
\label{deltagamma}
\end{eqnarray}
where $h(r)= 1/(1-r^{2})$.

Let us now rewrite  the bare level spectra 
$\epsilon_{n}= \omega_{c}\, e_{n}(g; \mu, d, r) $ 
as $\epsilon_{n}=   \epsilon_{n}^{\rm ren} + \delta \epsilon_{n}$.
The renormalized spectra $\epsilon_{n}^{\rm ren}$ are the portion 
consisting of $\omega_{c}^{\rm ren}$, 
$g^{\rm ren}\equiv \gamma_{1}^{\rm ren}/\omega_{c}^{\rm ren}$,
$d^{\rm ren}\equiv \Delta^{\rm ren}/\omega_{c}^{\rm ren}$, etc.
The associated counterterms $\delta \epsilon_{n}$, 
to be written as  $\delta \epsilon_{n} \equiv \lambda_{n}\,  \delta v/v^{\rm ren}$
with finite coefficients $ \lambda_{n}$, 
are expressed in terms of  $(\delta v, \delta \gamma_{1}, \delta \Delta)$ 
and are uniquely fixed once one specifies
$v^{\rm ren}$ (or, the divergence $\delta v$) by referring to a specific observable quantity;
see Appendix A for details. 
The dressed level spectra, rewritten as
\begin{equation}
\hat{\epsilon}_{n} = \epsilon_{n} + \Delta  \epsilon_{n}
= \epsilon_{n}^{\rm ren} + (\Delta  \epsilon_{n})^{\rm ren},
\end{equation}
thereby reveal observable self-energies 
$(\Delta  \epsilon_{n})^{\rm ren}\equiv \Delta  \epsilon_{n} + \delta \epsilon_{n}$,
with the divergences in $\Delta  \epsilon_{n}$ removed by $\delta \epsilon_{n}$.
Also the full CR energies $\epsilon_{\rm exc}^{n\leftarrow j}$  in Eq.~(\ref{fullEnj}) 
take a renormalized form, 
\begin{equation}
\epsilon_{\rm exc}^{n\leftarrow j} = \epsilon_{n}^{\rm ren}- \epsilon_{j}^{\rm ren} 
+ (\Delta \epsilon^{n,j})^{\rm ren},
\end{equation}
with corrections $(\Delta \epsilon^{n,j})^{\rm ren}$ free from divergences. 

Let us here define $v^{\rm ren}$ by referring to CR in the $-2 \leftarrow -3$ channel, 
as chosen experimentally~\cite{HJTS}. We fix $v^{\rm ren}$ so that 
$\epsilon^{-2\leftarrow -3}_{\rm exc}|^{n_{\rm f}=-3}_{(\mu, d,r) \rightarrow 0}$ 
takes a naive form $(\epsilon_{-2}^{\rm ren} -\epsilon_{-3}^{\rm ren})|_{X=0}$; 
i.e.,  $(\Delta \epsilon^{-2,-3})^{\rm ren}|_{X=0}=0$.
A neat way to handle this prescription is 
to replace ${\cal F}_{n}(z)$ in $\Delta \epsilon_{n}$  [of Eq.~(\ref{DEn_reg})] 
by 
 \begin{equation}
{\cal F}^{\rm ren}_{n}(z) 
= {\cal F}_{n}(z) - 2\lambda_{n}\, [\Gamma^{-2 \leftarrow -3}(z)]_{X=0},
\end{equation}
and cast $(\Delta \epsilon_{n})^{\rm ren}$ in the form
 \begin{eqnarray}
 (\Delta \epsilon_{n})^{\rm ren}  &=&\Omega_{n}
- \sum_{k} \nu [k]\,  \sum_{\bf p}v_{\bf p} \gamma_{\bf p}^{2}  |g^{nk}_{\bf p}|^2, 
\label{DEn_ren} \\
 \Omega_{n} &=&
{\textstyle{1\over{2}}}\sum_{\bf p}v_{\bf p}\, {\cal F}_{n}^{\rm ren}(z).
 \end{eqnarray}
Here $\Gamma^{-2 \leftarrow -3}(z)
 =\gamma^{2}_{\bf p}\, I_{z} / (\lambda_{-2} -\lambda_{-3} )$, with 
\begin{equation}
I_{z} =
-\sum_{k\le -3} (|g^{-2, k}_{\bf -p}|^{2}- |g^{-3, k}_{\bf p}|^{2})
- g^{-2,-2}_{\bf p}\, g^{-3,-3}_{\bf -p},
\end{equation}
is a kernel associated with $\Delta \epsilon^{-2,-3}$.
In this renormalized form, 
${\cal F}_{n}^{\rm ren}(z)$ are sizable only for $\ell |{\bf p}| \sim O(1)$ 
and vanish rapidly as $z = {1\over{2}}\ell^{2}{\bf p}^{2} \rightarrow \infty$; 
one can thus calculate many-body corrections $\Omega_{n}$ numerically 
without handling divergences.
Note  that we have carried out renormalization in such a way that 
${\cal F}^{\rm ren}_{n}(z)$ and hence $\Omega_{n}$ also share 
the same conjugation properties as 
$F_{n}(z)$ in Eq.~(\ref{eh_Fn}), i.e., 
 \begin{equation}
\Omega_{-n}= - \Omega_{n}|_{-X},\ \ 
\Omega_{n}^{K'} = \Omega_{n}^{K}|_{-\mu} =  - \Omega_{-n}^{K}|_{-d,-r}.
\end{equation}

Once the eigenmodes ${\bf b}_{n}$ are fixed numerically 
for a given set of  parameters $(v, \gamma_{1}, u, \Delta, r)$, 
one can now construct $g^{mn}_{\bf p}$, $(\epsilon_{n}, \lambda_{n})$,
and ${\cal F}^{\rm ren}_{n}(z)$ and calculate the spectra $\hat{\epsilon}_{n}$ 
and $\epsilon_{\rm exc}^{n\leftarrow j}$.
Table~I shows a list of $\Omega_{n}$ of our interest,   
in units of 
\begin{equation}
\tilde{V}_{c} = \sum_{\bf p}v_{\bf p}\gamma_{\bf p}^{2}
= {\alpha\over{\epsilon_{b}\ell}}  \sqrt{{\pi\over{2}}}
 \approx {70.3 \over{\epsilon_{b}}}\, 
\sqrt{B[{\rm T}]}\, {\rm meV}.
\end{equation}
For the PZM levels, $\Omega_{0}$ and $\Omega_{1}$ are practically linear in $\mu$ 
and are barely modified by $e$-$h$ breaking $(d,r)$; 
indeed, $(\Omega_{0}, \Omega_{1})|_{d=r=0} \approx (-0.6457,  -0.6120)\, \mu\, \tilde{V}_{c}$.
For other levels $e$-$h$ breaking is evident, 
$\Omega_{n} =-\Omega_{-n}|_{-X} \not= -\Omega_{-n}$.

\begin{table}
\caption{
Many-body corrections $\Omega_{n}$ in valley $K$ at $B=20$T.
Setting $\mu\rightarrow-\mu$ yields $\Omega_{n}$ in valley $K'$. 
}
\begin{ruledtabular}
\begin{tabular}{l }
{\ $\Omega_{4} \approx \tilde{V}_{c}\,  
(0.4758 + 0.0635 \mu - 0.0299 \mu^2)$} \\   
{\ $\Omega_{3} \approx \tilde{V}_{c}\,  
(0.4363 + 0.0884 \mu + 0.157 \mu^2)$} \\   
{\ $\Omega_{2} \approx \tilde{V}_{c}\,  
(0.3539 + 0.146 \mu + 0.663 \mu^2)$} \\   
{\ $\Omega_{0} \approx \tilde{V}_{c}\,  
(\ \  0\ \ \ \ \ -0.6446 \mu - 0.0170 \mu^2)$} \\  
{\ $\Omega_{1} \approx \tilde{V}_{c}\,  
(-0.00339 - 0.6109 \mu - 0.0317 \mu^2)$} \\   
  {$\Omega_{-2}\! \approx \tilde{V}_{c}\,  
(-0.4136 + 0.222 \mu - 0.641 \mu^2)$} \\   
{$\Omega_{-3}\! \approx \tilde{V}_{c}\,  
(-0.4995 + 0.114 \mu - 0.166 \mu^2)$} \\   
{$\Omega_{-4}\! \approx \tilde{V}_{c}\,  
(-0.5441 + 0.0719 \mu + 0.00489 \mu^2)$} \\   
\end{tabular}
 \vskip0.1cm
\end{ruledtabular}
\end{table}

\section{Orbital mixing and Cyclotron resonance within the PZM sector}

In bilayer graphene the LLL consists of $2_{\rm spin}\times2_{\rm valley}=4$ sets 
of nearly degenerate $n=\{0,1\}$ levels.
For the empty PZM sector [of a given (spin,  valley)] at $N_{\rm f}=0$ 
one can explicitly write down the renormalized spectra by substituting  
$\nu [k] = -{1\over{2}} (\delta_{k,0}+ \delta_{k,1})$ into Eq.~(\ref{DEn_ren}),
\begin{eqnarray}
\hat{\epsilon}_{0}\!\!\!
&\stackrel{N_{\rm f}=0}{=}&\!\!\! -\omega_{c}\mu + \Omega_{0}
+ {\textstyle{1\over{2}}} ( 1+  {\textstyle{1\over{2}}}\, c_{1}^2 )\, \tilde{V}_{c},
\nonumber\\
\hat{\epsilon}_{1}\!\!\!
&\stackrel{N_{\rm f}=0}{=}&\!\!\! 
-\omega_{c} (\mu -\kappa)
+\Omega_{1}
+  {\textstyle{1\over{2}}} ( 1+ {\textstyle{1\over{2}}}\, c_{1}^2 
-  {\textstyle{1\over{4}}}\, C)  \tilde{V}_{c},
\label{Spec_pzm_empty}
\end{eqnarray}
with $C= (4 - 3  c_{1}^{2})\, c_{1}^{2}$. 
Here and from now on all quantities refer to renormalized ones. 
In this section we use  $\hat{\epsilon}_{n}$ to denote  
the $N_{\rm f}=0$ spectra, $\hat{\epsilon}_{n}= \hat{\epsilon}_{n}|^{N_{\rm f}=0}$.

In Eq.~(\ref{Spec_pzm_empty}), 
the first terms $\ni (\mu, \kappa, \Omega_{0},\Omega_{1})$ are odd in breaking $X = (\mu,d,r)$ 
while the last terms $\propto \tilde{V}_{c}$ are even in  $X$.
For  $X =0$, only the latter remain.
The orbital degeneracy is therefore lifted by quantum corrections even for zero breaking $X=0$, 
giving rise to the orbital Lamb shift~\cite{KS_Ls},
with $\hat{\epsilon}_{1}$ lower than  $\hat{\epsilon}_{0}$ by 
\begin{equation}
\epsilon_{\rm Ls} \equiv {\textstyle{1\over{8}} }\, C\,  \tilde{V}_{c}.
\end{equation}
Numerically, at $B=20$\,T, 
$C\approx 1.21 + 0.016 \mu + 0.04 \mu^2$,
$(\hat{\epsilon}_{0}, \hat{\epsilon}_{1})|_{X=0} \approx (0.72, 0.57) \tilde{V}_{c}$ and
$\epsilon_{\rm Ls} \approx 0.15  \tilde{V}_{c}$, or 
$\epsilon_{\rm Ls} \approx  8.3$\,meV
for a choice of $\tilde{V}_{c}/\omega_{c} =0.4$ or $\epsilon_{b} \approx 5.7$.
The dressed spectra $\hat{\epsilon}_{n}$ considerably deviate 
from the one-body spectra $\epsilon_{n}$, 
as shown in Fig.~1 for $\tilde{V}_{c}/\omega_{c} =0.4$ and $B=20$T at bias $u=10$ meV.

\begin{figure}[tbp]
\begin{center}
\includegraphics[scale=.63]{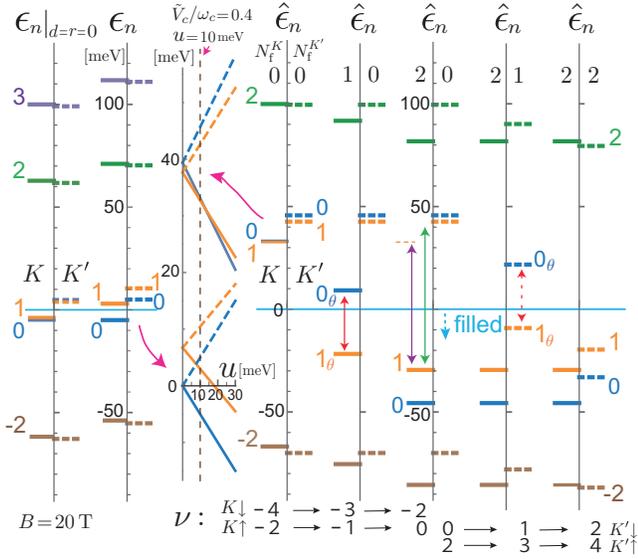}
\end{center}
\caption{
Landau-level spectra in valleys $(K,K')$ at bias $u=$\,10\,meV and $B=20$\,T;
spin splitting is suppressed. 
(left)~One-body spectra $\epsilon_{n}|_{d=r=0}$ and $\epsilon_{n}$. 
$e$-$h$ breaking $(d,r)$ shifts the $e$-$h$ symmetric spectra
$\epsilon_{n}|_{d=r=0}$ upwards appreciably, except for $\epsilon_{0}$. 
(right)~Coulomb-corrected spectra $\hat{\epsilon}_{n}$ 
for $\tilde{V}_{c}/\omega_{c}=0.4$. 
Level spectra evolve with filling $(N_{\rm f}^{K}, N_{\rm f}^{K'})$ of each valley and spin.
When bias $u$ dominates over spin splitting $\mu_{\rm Z}$, 
filling of the LLL will start with valley $K$ with spin 
$\downarrow$ at $\nu=-4$, as illustrated in the figure. 
Green arrows denote possible $\nu=0$ gaps, purple arrows $\nu=\mp 2$ gaps 
and red arrows orbital gaps at odd-integer filling.
}
\end{figure}

Let us write, for $X\not=0$, the full $(0,1)$ shift as
\begin{equation}
\hat{\epsilon}_{0}-\hat{\epsilon}_{1} 
=\epsilon_{\rm Ls} + \Omega_{0} -\Omega_{1} -\omega_{c}\kappa 
\equiv (1 -\xi)\, \epsilon_{\rm Ls}.
\end{equation}
Numerically, at $B=20$T,
\begin{equation}
\xi \approx -0.023 + 0.22 \mu 
  + (0.325 + 1.7 \mu)/(\tilde{V}_{c}/\omega_{c}).
\end{equation}
Obviously, $\xi > 0$ for $\mu \ge 0$.
The $O(X)$ term in $\hat{\epsilon}_{1}$,
$\kappa \approx 2\mu/(g^2 +1) + O(d, r)$, 
thus tends to reduce the orbital Lamb shift, e.g., 
$(1- \xi)\, \epsilon_{\rm Ls} \approx (0.53, 0.21, -0.11) \epsilon_{\rm Ls}$
for  $u=(-20, 0, 20)$ meV at $B=20$T. 
At the same time, $\kappa$ makes $\hat{\epsilon}_{1}$
slightly less dependent on $\mu$ than $\hat{\epsilon}_{0}$. 
As a result, the nearly degenerate PZM levels 
$(\hat{\epsilon}_{0},\hat{\epsilon}_{1})$, being quasi linear in $\mu$,
necessarily have a crossing or level inversion in either valley (per spin) 
as bias $u$ is varied.

 \begin{figure}[tbp]
\begin{center}
\includegraphics[scale=.68]{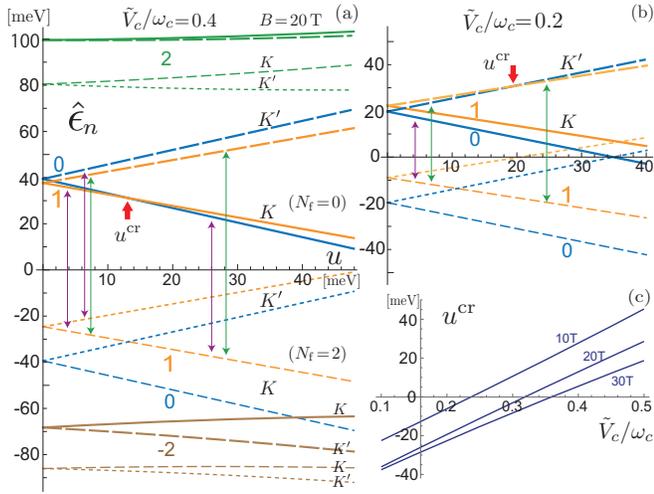}
\end{center}
\caption{
Level spectra at $B=$20\,T 
for (a) $\tilde{V}_{c}/\omega_{c}=0.4$ and  (b) $\tilde{V}_{c}/\omega_{c}=0.2$.
Bold (solid/long-dashed) curves refer to the spectra (in valley $K/K'$) at $N_{\rm f}=0$
[with empty $n=(0,1)$ levels], and thin (dashed/dotted) curves to those at $N_{\rm f}=2$ 
[with filled $(0,1)$ levels].
Level inversion takes place at $N_{\rm f}=0$ across $u=u^{\rm cr}$;
 $u^{\rm cr}\approx$13\,meV in (a) and $u^{\rm cr}\approx$ -19\,meV in (b).
The $\nu=0$ band gaps (green arrows) increase with bias $u$ 
while possible $\nu=\mp 2$ gaps (purple arrows) barely change.
(c) $u^{\rm cr}$ as a function of $\tilde{V}_{c}/\omega_{c}$ at $B=$(10,20,30)\,T.
}
\end{figure}

These features are indeed seen from Fig.~2(a), which shows 
how such $N_{\rm f}=0$ level spectra (per spin) change with bias $u$
for $\tilde{V}_{c}/\omega_{c} = 0.4$;
solid curves refer to valley $K$ and long-dashed curves to $K'$; 
$\hat{\epsilon}_{n}|^{K'}= \hat{\epsilon}_{n}|^{K}_{-\mu}$.
Level inversion takes place across $\xi =1$, or numerically, 
across $\mu^{\rm cr}\approx 0.0475$  or $u^{\rm cr}\approx 13$ meV.
Such a critical bias $u^{\rm cr}$ gets smaller 
as $\tilde{V}_{c}$ is made weaker and for higher $B$, 
as seen from Figs.~2(b) and 2(c); 
$u^{\rm cr}>0$ lies in valley $K$ while $u^{\rm cr}< 0$ lies in valley $K'$.
Level inversion is also present for $d=r=0$, with
$u^{\rm cr} \approx 62$meV at $B=20$T.

When the PZM sector is gradually filled with electrons, 
those level spectra come down (via exchange interaction) 
and, when the sector is filled up, they turn into the ones, 
depicted with thin dashed and dotted curves in Fig.~2. 
These $N_{\rm f}=2$ spectra 
are $e$-$h$ conjugate to the $N_{\rm f}=0$ spectra, 
\begin{equation}
(\hat{\epsilon}_{0}, \hat{\epsilon}_{1})|^{N_{\rm f}=2} 
= (-\hat{\epsilon}_{0}, -\hat{\epsilon}_{1})|^{N_{\rm f}=0}_{-X},
\end{equation}
and are given by 
\begin{eqnarray}
\hat{\epsilon}_{0}|^{N_{\rm f}=2} \!
&=&\! -\omega_{c}\mu + \Omega_{0} 
-{\textstyle{1\over{2}}} ( 1+  {\textstyle{1\over{2}}}\, c_{1}^2 )\, \tilde{V}_{c},
\nonumber\\
\hat{\epsilon}_{1}|^{N_{\rm f}=2}\!
&=&\! 
-\omega_{c} (\mu - \kappa)
+\Omega_{1}
-  {\textstyle{1\over{2}}} ( 1+ {\textstyle{1\over{2}}} c_{1}^2 
-  {\textstyle{1\over{4}}} C)  \tilde{V}_{c}. \ \
\label{Spec_pzm_full}
\end{eqnarray}
At $N_{\rm f}=2$, orbital shift 
$(\hat{\epsilon}_{0}-\hat{\epsilon}_{1})|^{N_{\rm f}=2} 
=  -(1+\xi)\, \epsilon_{\rm Ls}<0$
 is reversed in the $(0,1)$ layout and is considerably enhanced by $\xi$,
as is clear from Fig.~2.
Incidentally, for $(d, r) =0$, these conjugate $N_{\rm f}=(0,2)$ spectra
look symmetric about the $u$ axis.
In this sense, $e$-$h$ breaking $(d,r)$ is seen as an apparent asymmetry in Fig.~2. 
[Actually, asymmetry is particularly sizable for the $n=1$ level;
with $R_{n}\equiv \hat{\epsilon}_{n}/(\hat{\epsilon}_{n}|_{d=r=0})$, 
one finds $R_{1} \approx 1.2$,
$R_{0} \approx 1$ 
and $R_{n} \approx 1.07$
($n=\pm2 \sim \pm 6$) for $\tilde{V}_{c}/\omega_{c} =0.4$ and $u \sim 0$.]

Let us now consider what will happen when we pass from $N_{\rm f}=0$ to $N_{\rm f}=2$ 
in valley $K$ with bias $u$ kept fixed in the range $0<u < u^{\rm cr}$. 
It is the $n=1$ level that starts to be filled, getting lower in energy.
The $n=0$ level also follows but, when $N_{\rm f}=2$ is reached, 
it gets even lower than the $n=1$ level. 
This signals a level crossing or instability with filling, 
which actually is avoided via mixing of $n=\{0,1\}$ levels~\cite{KS_Ls}. 
For $u > u^{\rm cr}$  no such mixing takes place.  
Orbital mixing is therefore triggered by level inversion in the PZM spectra. 
(i) When $u^{\rm cr}\ge 0$, orbital mixing arises in both valleys for $0 \le u < u^{\rm cr}$
and in one valley ($K'$) for $u > u^{\rm cr}$.
(ii) When $u^{\rm cr} < 0$, mixing takes place only in one valley ($K'$) for $u > |u^{\rm cr}|$, 
as seen from Fig.~2(b). 
(Here we suppose no level inversion in the $N_{\rm f}=2$ spectra 
and take, e.g., $u \lesssim 50$\,meV.)

To see how level mixing proceeds let us rotate $\hat{\psi} =(\psi^{0}, \psi^{1})^{\rm t}$ 
to $\hat{\Phi} =(\Phi^{0}, \Phi^{1})^{\rm t} = U \hat{\psi}$ 
in the orbital space,
\begin{equation}
\hat{\psi} = U \hat{\Phi} =  \left(
\begin{array}{cc}
c_{\theta} & -s_{\theta} \\
s_{\theta}&c_{\theta}\\
\end{array}
\right)\left(
\begin{array}{c}
\Phi^{0}\\ 
\Phi^{1} \\
\end{array}\right),
\end{equation}
where $c_{\theta}\equiv \cos (\theta/2)$ and $s_{\theta}\equiv \sin (\theta/2)$;
we refer to the levels associated with $\Phi^{n}$ as 
$n=(0_{\theta}, 1_{\theta})$.
We also define the rotated charges $S_{\bf p}^{mn}$ by 
$g^{mn}_{\bf -p}\, R^{mn}_{\bf p} = G^{mn}_{\bf -p}\, S_{\bf p}^{mn}$,
where $G^{mn}_{\bf -p} = (U^{\dag} g_{\bf -p} U)^{mn}$ for $n,m \in (0,1)$,
$G^{mn}_{\bf -p} = (U^{\dag} g_{\bf -p})^{mn}$ 
for $m \in (0,1)$ and $n \notin (0,1)$, etc.;
$S_{\bf -p}^{mn}$ are written with 
$(\Phi^{0},\Phi^{1})$ and $\psi^{j}$ with $j\notin (0,1)$.

The PZM sector for  $0\le N_{\rm f} \le 2$ is now governed 
by the one-body Hamiltonian [with the $N_{\rm f}=0$ spectra 
$(\hat{\epsilon}_{0},\hat{\epsilon}_{1})$ for $(\psi^{0},\psi^{1})$]  
plus Coulomb interaction $V^{\rm pzm}$
[with $g_{\bf p} R_{\bf -p} \rightarrow G_{\bf p} S_{\bf -p}$ 
in Eq.~(\ref{VCoul})] acting within this sector. 
One can readily diagonalize it using the Hartree-Fock approximation: 
The rotated levels $(0_{\theta}, 1_{\theta})$ have the spectra~\cite{KS_Ls},
\begin{eqnarray}
\hat{\epsilon}_{0_{\theta}}&=& \hat{\epsilon}_{+} 
+\hat{\epsilon}_{-}  \cos \theta
+ N_{0}\, E^{00}(\theta) + N_{1}\, E^{01}(\theta),
\nonumber\\
\hat{\epsilon}_{1_{\theta}}&=& \hat{\epsilon}_{+}  
-\hat{\epsilon}_{-} \cos \theta 
+ N_{1}\, E^{11}(\theta) + N_{0}\, E^{01}(\theta),
\label{EPZM-theta}
\end{eqnarray}
where $\hat{\epsilon}_{+} = {1\over{2}}\, (\hat{\epsilon}_{0} + \hat{\epsilon}_{1})$,
$\hat{\epsilon}_{-} ={1\over{2}}\, (1-\xi)\, \epsilon_{\rm Ls}$ and 
\begin{eqnarray}
E^{00}(\theta) &=&  - \tilde{V}_{c} \,  \left [1- \textstyle{1\over{16}}C\, (1- \cos \theta)^2 \right],
\nonumber\\
E^{11}(\theta) &=& - \tilde{V}_{c} \,  \left [1- \textstyle{1\over{16}}C\, (1+ \cos \theta)^2\right],
\nonumber\\
E^{01}(\theta) &=& 
- \tilde{V}_{c} \, \left [\textstyle{1\over{2}}c_{1}^{2} - {1\over{16}}\, C \{1-(\cos \theta)^2\} \right];
\end{eqnarray}
$(N_{0}, N_{1})$ denote the filling fractions of the $(0_{\theta}, 1_{\theta})$ levels,
with $0 \le N_{0}\le 1$, etc. 
Note that $(\hat{\epsilon}_{0_{\theta}}, \hat{\epsilon}_{1_{\theta}})$
enjoy the reciprocal relation 
\begin{equation}
\hat{\epsilon}_{1_{\theta}}
= \hat{\epsilon}_{0_{\theta}}|_{N_{0} \leftrightarrow N_{1},  \theta\rightarrow \pi - \theta}.
\end{equation}

Diagonalization of the PZM spectra is achieved for $\theta$ that obeys the relation
\begin{eqnarray}
&&\partial_{\theta}\left [\epsilon_{0} (\theta) 
+  \textstyle{1\over{2}} \, \{  N_{0} \, E^{00}(\theta) 
  - N_{1}\, {E}^{11}(\theta) \} \right] 
\nonumber\\
&\propto& \sin \theta
\left [N_{0} +N_{1}   -(1-\xi)+ (N_{1}-N_{0}) \cos \theta  \right]=0,\ \ \ \ \ 
\label{fixTheta}
\end{eqnarray}
i.e., $\sin \theta=0$ or 
$\cos \theta =(1 - \xi+  N_{0} - N_{1})  /(N_{1} -N_{0})$,
where
$\epsilon_{0}(\theta) 
= c_{\theta}^2\, \hat{\epsilon}_{0}+ s_{\theta}^2\, \hat{\epsilon}_{1}
= \hat{\epsilon}_{+} +\hat{\epsilon}_{-}  \cos \theta$.

 \begin{figure}[bpt]
\begin{center}
\includegraphics[scale=1.0]{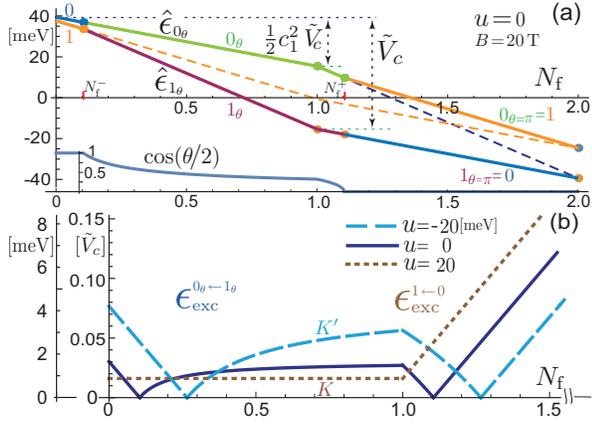}
\end{center}
\vskip-.7cm
\caption{Orbital mixing. (a) PZM spectra 
$(\hat{\epsilon}_{0_{\theta}},  \hat{\epsilon}_{1_{\theta}})$ 
over the range $0\le N_{\rm f} \le 2$ for $\tilde{V}_{c}/\omega_{c}=0.4$ at $u=0$ and $B=20$\,T. 
Dashed lines refer to the level spectra with no rotation $\theta=0$. 
(b) Cyclotron resonance within the PZM sector at bias $u=(0,\pm 20)$\,meV.
}
\end{figure}

Figure~3(a) depicts how the $(\hat{\epsilon}_{0_{\theta}},\hat{\epsilon}_{1_{\theta}})$ spectra 
change with filling factor $N_{\rm f}=N_{1}+ N_{0}$  
for $u=0 <u^{\rm cr}$ and $B=20$\,T.
For $u<u^{\rm cr}$ critical changes arise at three points 
$(N_{\rm f}^{-},1, N_{\rm f}^{+} =1+ N_{\rm f}^{-})$
with $N_{\rm f}^{-} = {1\over{2}} (1-\xi) < 0.5$, and  
orbital mixing takes place in the range
$(N_{\rm f}^{-}, N_{\rm f}^{+})$:
(i)~The $1_{\theta}$ level first gets filled. 
$\theta$ deviates from zero at $N_{\rm f} (=N_{1}) = N_{\rm f}^{-}$, 
and changes as
\begin{equation}
\cos \theta =(1-\xi)/N_{\rm f}  -1,
\end{equation}
until $\cos \theta =-\xi <0$ is reached at $N_{\rm f}=1$.
(ii) Filling of the $0_{\theta}$ level then starts,  and 
$\theta$ changes according to 
\begin{equation}
\cos \theta =(1 -\xi -N_{\rm f})/(2-N_{\rm f} ).
\end{equation}
$\cos \theta$ attains $-1$ or $\theta =\pi$ at $N_{\rm f}=N_{\rm f}^{+}$ 
and ceases to change.
Via mixing the $\{0,1\}$ levels are interchanged,
\begin{equation}
(\psi^{0}, \psi^{1}))|_{\theta=0}  
\rightarrow (\Phi^{0}, \Phi^{1})|_{\theta=\pi} = (\psi^{1}, -\psi^{0}).
\end{equation}

Actually, noting Eq.~(\ref{fixTheta}), one can simplify the level spectra
when $\theta$ is moving, i.e., for
$N_{\rm f}^{-} \le N_{\rm f} \le N_{\rm f}^{+}$,
\begin{eqnarray}
\hat{\epsilon}_{0_{\theta}}&=& \hat{\epsilon}_{0}
 -\tilde{V}_{c}\, ( N_{0} +{\textstyle {1\over{2}} } c_{1}^{2} N_{1}  ),
 \nonumber\\
\hat{\epsilon}_{1_{\theta}} &=& \textstyle
\hat{\epsilon}_{0} -\tilde{V}_{c}\, ( N_{1} + {\textstyle {1\over{2}} } c_{1}^{2} N_{0}  ).
\end{eqnarray}
The $N_{\rm f}=1$ spectra 
$( \hat{\epsilon}_{0_{\theta}}, \hat{\epsilon}_{1_{\theta}})|^{N_{\rm f}=1}
= (\hat{\epsilon}_{0} -{\textstyle {1\over{2}} } c_{1}^{2}  \tilde{V}_{c}, 
\hat{\epsilon}_{0} - \tilde{V}_{c})$
then obey the relation
\begin{equation}
( \hat{\epsilon}_{0_{\theta}}, \hat{\epsilon}_{1_{\theta}})|^{N_{\rm f}=1}
= (-\hat{\epsilon}_{1_{\theta}}, - \hat{\epsilon}_{0_{\theta}})|^{N_{\rm f}=1}_{-X},
\end{equation}
which implies that, via $e$-$h$ conjugation, 
the filled $1_{\theta}$ level turns into the filled $0_{\theta}$ within a valley. 
The small gap $\hat{\epsilon}_{0} - \hat{\epsilon}_{1} 
=(1- \xi)\, \epsilon_{\rm Ls}$ at $N_{\rm f}=0$ develops into a sizable orbital gap 
$(1 - {\textstyle {1\over{2}} } c_{1}^{2} )\, \tilde{V}_{c}$ 
at $N_{\rm f}=1$, as seen from Fig.~3(a).

On the other hand, for $u > u^{\rm cr }$, no such mixing arises, but 
$(\hat{\epsilon}_{0}, \hat{\epsilon}_{1})$ show similar behavior with filling.
At $N_{\rm f}=1$, 
$( \hat{\epsilon}_{0}, \hat{\epsilon}_{1})|^{N_{\rm f}=1}
= (\hat{\epsilon}_{0} -\tilde{V}_{c}, 
\hat{\epsilon}_{1}-{\textstyle {1\over{2}} } c_{1}^{2}\,  \tilde{V}_{c})$ 
correctly obey the relation in Eq.~(\ref{Nfone_NoRotation}) and have again a sizable gap.

In bilayer graphene the empty LLL (at $\nu=-4$) consists of 
four sets of empty PZM sectors. 
Let us now consider how the LLL is filled 
over the range $-4 \le \nu\le 4$.
As an illustration we focus on some typical cases 
in which the eightfold degeneracy of the LLL is fully lifted. 
Suppose first that bias $u$ is chosen so that the valley gap~$\sim O(u)$ 
dominates over the spin gap $\mu_{\rm Z}$.
Filling of the LLL will then starts in valley $K$ with a spin component of the lowest energy, 
and each \{valley, spin\} set of PZM sectors will be filled in the following order  
\begin{equation}
\stackrel{\nu=-4}{\rightarrow}
 \{K\downarrow\} \stackrel{\nu=-2}{\rightarrow} \{K\uparrow\} 
\stackrel{\nu=0}{\rightarrow} \{K'\downarrow\}  
\stackrel{\nu=2}{\rightarrow} \{K'\uparrow\}. 
\label{fillingLLL_valley}
\end{equation}
Figure 1 illustrates such a sequence of level spectra $\{ \hat{\epsilon}_{n}\}$ for $u=10{\rm meV}<u^{\rm cr}$.
There the total filling factor $\nu$ increases with each valley-filling factor 
$\{N_{\rm f}^{K\downarrow}, N_{\rm f}^{K\uparrow}, \cdots\}$,
and a  sizable band gap emerges at each integer filling:
(i)~The $\nu=\pm 3$ and $\nu=\pm1$ gaps are orbital gaps (in red) of width 
$(1-  {1\over{2}}c_{1}^{2})\, \tilde{V}_{c}\ (\sim 0.56 \tilde{V}_{c})$.
(ii)~The $\nu=\mp 2$ gaps are enhanced spin gaps  (in purple),
$\hat{\epsilon}_{1}|^{N_{\rm f}=0}_{-\mu} - \hat{\epsilon}_{0}|^{N_{\rm f}=2} 
\approx (1+  {1\over{2}}c_{1}^{2} - {1\over{4}} C)\, \tilde{V}_{c} \ (\sim 1.13 \tilde{V}_{c})$,
associated with full filling of one valley.
(iii)~Most prominent is the $\nu=0$ gap, which is a valley (+ orbital) gap (in green), 
$\hat{\epsilon}_{1}|^{K';N_{\rm f}=0}_{-\mu} - \hat{\epsilon}_{1}|^{K;N_{\rm f}=2}$
$\sim u +   (1+  {1\over{2}}c_{1}^{2} - {1\over{4}} C)\, \tilde{V}_{c}$.
As seen from Figs.~2(a) and 2(b), 
the $\nu=0$ gap and $\nu=\pm 2$ gaps are essentially the same at zero bias $u=0$, 
but the former increases rapidly with $u$ while the latter barely change.

On the other hand, when the spin gap dominates over the valley gap in high field $B$, 
i.e., $\mu_{Z} \gg u \sim +0$, filling of the LLL will proceed in the following order 
 \begin{equation}
\stackrel{\nu=-4}{\rightarrow}
 \{K\downarrow\} \stackrel{\nu=-2}{\rightarrow} \{K'\downarrow\} 
\stackrel{\nu=0}{\rightarrow} \{K\uparrow\}  
\stackrel{\nu=2}{\rightarrow} \{K'\uparrow\},
\label{fillingLLL_spin}
\end{equation}
and the $\nu=0$ gap $\propto \mu_{Z} + (1+  {1\over{2}}c_{1}^{2} - {1\over{4}} C)\, \tilde{V}_{c}$ 
will open as an enhanced spin gap. 
Two such filling sequences (\ref{fillingLLL_valley}) and (\ref{fillingLLL_spin}) 
appear consistent with part of the ($u$-dependent)  layer-charge pattern 
observed in a recent capacitance measurement~\cite{KFLH}.

Let us next consider CR supported by the PZM sector (per spin and valley).
Consider a time-dependent uniform electric potential 
$A(t)= A = A_{x}(t) + iA_{y}(t)$, coupled to the PZM sector 
via the currents $R^{01}$ and $R^{1,\pm 2}$, 
with the Hamiltonian 
 \begin{equation}
 H_{A} = ev\,  \zeta_{1}A_{\theta}\,  S_{\bf p=0}^{01}  
+ev\zeta_{n}  A\, 
( c_{\theta} S_{\bf 0}^{1n}+ s_{\theta} S_{\bf 0}^{0n})
+ \cdots,
\label{H_Aelectric}
\end{equation}
where $A_{\theta}= c_{\theta}^2A -s_{\theta}^2 A^{\dag}$,
$\zeta_{1} = b_{1}- r\, c'_{1} \approx -\kappa\, c_{1}$ and 
$\zeta_{\pm 2} \approx b'_{\pm 2}\, c'_{1} + b_{\pm 2}\, c_{1}$.
Accordingly, CR takes place in channels  
$0_{\theta} \leftarrow 1_{\theta}$, 
 $2 \leftarrow (1_{\theta}, 0_{\theta})$,
 $(1_{\theta}, 0_{\theta}) \leftarrow -2$, etc.

Notably, intra PZM resonance is possible~\cite{BCNM}. 
When orbital mixing is present ($u<u^{\rm cr}$),  
CR arises in the $0_{\theta} \leftarrow 1_{\theta}$ channel,
with excitation energy
\begin{eqnarray} 
\epsilon_{\rm exc}^{0_{\theta}\leftarrow 1_{\theta}}\!
&=& \hat{\epsilon}_{0_{\theta}} - \hat{\epsilon}_{1_{\theta}}\! - (N_{1}-N_{0})\, {\cal A}_{01},
\nonumber\\
&=&\! \epsilon_{\rm Ls}\, \big\{   (1- \xi)x+ N_{0}\, q[x] 
 - N_{1}\, q[-x] \big\},
 \nonumber\\
 q[x] &=&  (1/2)(1+x)(1-3x), 
\end{eqnarray} 
where 
$x\equiv \cos \theta$, $\epsilon_{\rm Ls} =  {1\over{8}} C\,\tilde{V}_{c}$
and
\begin{equation}
{\cal A}_{01}\! =  \sum_{\bf p}\! v_{\bf p} \gamma_{\bf p}^{2} 
G^{00}_{\bf -p} G^{11}_{\bf p}
= (1 - \textstyle{1\over{2}} c_{1}^2)\, \tilde{V}_{c} 
- \textstyle{1\over{2}}\epsilon_{\rm Ls}\, \sin^{2} \theta.
\end{equation}

On the other hand,  when orbital mixing is absent $(u>~u^{\rm cr})$,
the $1\leftarrow 0$ channel is activated, 
with 
\begin{equation}
\epsilon_{\rm exc}^{1\leftarrow 0}
=  \textstyle   \hat{\epsilon}_{1} - \hat{\epsilon}_{0}
+  {1\over{4}} C\tilde{V}_{c}\, N_{1}
= \epsilon_{\rm Ls}\, \{ \xi -1 + 2N_{1} \}.
\end{equation}
Interestingly, for $0 < N_{\rm f}(=N_{0}) \le 1$, there is no correction to CR
since $|g^{10}_{\bf p}|^2 -|g^{00}_{\bf p}|^2 + g^{11}_{\bf p} g^{00}_{\bf -p} =0$
holds; see Eq.~(\ref{gmnPZM}).

Figure~3(b) shows how the excitation spectra evolve with filling $N_{\rm f}$
at bias $u=0$,  $u=-20$ meV$ <u^{\rm cr}$ (thus in valley $K'$) 
and $u=20$ meV $>u^{\rm cr}$.
While a sizable level gap opens around $N_{\rm f} \sim 1$, 
it is almost cancelled by Coulombic attraction, 
leaving $\epsilon_{\rm exc}^{0_{\theta}\leftarrow 1_{\theta}}$
and $\epsilon_{\rm exc}^{1\leftarrow 0}$ of magnitude of the Lamb shift 
$\epsilon_{\rm Ls}  \sim O(0.15\, \tilde{V}_{c})$ or even smaller.

Such an intra-PZM channel of CR is activated 
when an orbital band gap develops around $\nu \sim (\pm 3, \pm1)$.
The coupling $\zeta_{1}\approx -\kappa\, c_{1}$ to the $S^{01}$ charge, 
however, is weak. 
It will therefore be a challenge to detect CR within the LLL in experiment. 
On the contrary, CR from or into the LLL serves as a practical probe into the LLL, 
as discussed in the next section.

\section{ interband cyclotron resonance }


\begin{figure*}[tbp]
\begin{center}
\includegraphics[scale=.95]{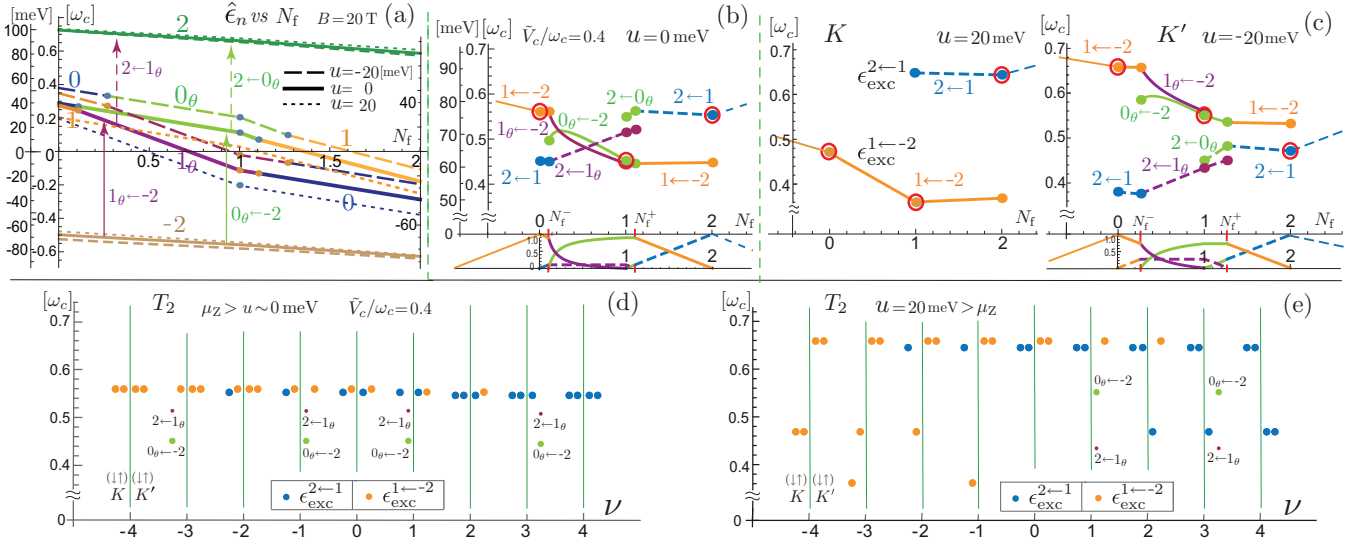}
\end{center}
\vskip-.7cm
\caption{Cyclotron resonance $T_{2} = \{ 2\leftarrow {\rm PZM}, {\rm PZM} \leftarrow-2 \}$,
with the choice $\tilde{V}_{c}/\omega_{c}=0.4$ at $B$=20\,T.
(a)~Evolution of the level spectra with filling of the PZM sector ($N_{\rm f}=0 \rightarrow 2$).  
Vertical arrows denote the associated CR channels at zero bias $u=0$.
(b)~Resonance spectra, in units of $\omega_{c}$ and meV, over the range 
$0\lesssim N_{\rm f} \lesssim 2$ at zero bias $u=0$;
also shown in scale $(0 \sim 1.0)$ is the relative strength of each resonance signal.
Red circles indicate the most prominent signal at each integer filling.
(c)~Resonance spectra at bias $u=20$\,meV $> u^{\rm cr}(\approx 13 {\rm meV})$ 
in valley $K$, and at  $u|^{K'}=-20$\,meV $< u^{\rm cr}$ in valley $K'$. 
(d)~Evolution of the full $T_{2}$ spectra in four \{spin,valley\} channels with filling of the LLL 
according to the spin-dominant filling sequence of Eq.~(\ref{fillingLLL_spin}).
(e)~Evolution of the $T_{2}$ spectra according to 
the valley-dominant filling sequence of Eq.~(\ref{fillingLLL_valley}).
}
\end{figure*}


\subsection{$T_{2}=\{2\leftarrow$ PZM, PZM$\leftarrow -2$\} }

Figure~4(a) shows how the level spectra evolve 
as the PZM sector is gradually filled over
$0~\le N_{\rm f} \le~2$ (per spin and valley).  
The PZM spectra are shifted almost uniformly with bias $u$ 
while the $n=\pm 2$ spectra are barely affected.  
As a result, the associated $T_{2}$ channels of CR sensitively depend on $u$.

For $0\le u < u^{\rm cr}$ orbital mixing is present in both valleys 
and CR arises in four channels, $2 \leftarrow (1_{\theta}, 0_{\theta})$ and 
$(1_{\theta}, 0_{\theta}) \leftarrow -2$ over the range 
$N_{\rm f}^{-} < N_{\rm f} < N_{\rm f}^{+}$;
see Appendix  B for details of the CR spectra.
The resonance spectra change in composition and strength 
in a characteristic way with filling $N_{\rm f}$, 
as depicted in Fig.~4(b) 
for $\tilde{V}_{c}/\omega_{c}=0.4$ at  $u=0$ and $B=$\,20\,T.
The strength of each response (function) is proportional to 
$(\nu_{\rm i}-\nu_{\rm f})\times w_{\rm mix}$, 
i.e, the filling-factor difference between the initial and final levels 
$\times$  mixing weight $w_{\rm mix}$ read from $H_{A}$ in Eq.~(\ref{H_Aelectric}).
In Fig.~4(b) we also plot such relative weights,  
$\{N_{1} c_{\theta}^2, N_{0} s_{\theta}^2\}$
for $2\leftarrow \{1_{\theta}, 0_{\theta}\}$ and
$\{(1-N_{1})c_{\theta}^2, (1-N_{0})s_{\theta}^2\}$
for $ \{1_{\theta}, 0_{\theta}\} \leftarrow -2$.
In the spectra red circles  indicate the most prominent signal at each integer filling.

The $1\leftarrow -2$ and $2\leftarrow 1$ channels remain active 
even for $-1< N_{\rm f}<0$ and $2< N_{\rm f}<3$, respectively, 
i.e., when either the $n=-2$ or  $n=2$ level is partially filled. 
The associated CR spectra slightly rise there, because Coulombic attraction diminishes 
as $N_{\rm f} \rightarrow -1$ or  $N_{\rm f} \rightarrow 3$; 
the CR signals themselves also vanish in this limit.

When bias $u$ is increased, e.g., to $u=20\, {\rm meV} > u^{\rm cr }$, 
orbital mixing disappears in valley $K$, with only the $1\leftarrow -2$ and 
$2\leftarrow 1$ channels of CR activated, 
while orbital mixing continues in valley $K'$, 
as depicted in Fig.~4(c).

From these spectra one can visualize 
how the $2_{\rm spin}\times 2_{\rm valley}=4$ channels (or more) of $T_{2}$ 
compete as the LLL is filled over the range $-4 \le \nu \le 4$ under a given bias $u$.
Figure~4(e) illustrates such resonance spectra of $u=20$\,meV at each integer filling, 
with the valley-dominant filling sequence $(u\gg \mu_{\rm Z})$ 
of Eq.~(\ref{fillingLLL_valley}) assumed.
Also Fig.~4(d) shows the $T_{2}$ spectra, with the spin-dominant filling sequence
($\mu_{\rm Z}> u \sim +0)$ of Eq.~(\ref{fillingLLL_spin}) assumed.
It is clear that, with increasing bias $u$, the spectra get split in the valley.
Remarkably, the resonance spectra look nearly $e$-$h$ symmetric, i.e., symmetric in $\nu$.
An asymmetry is apparent at odd integers $\nu=(\pm1, \pm3)$ in Fig.~4(e), 
where orbital mixing only arises in valley $K'$.

It is illuminating here to look into some numbers.
From Figs.~4(a)$\sim$(c) one can read off the following set of level gaps 
$\delta\hat{\epsilon}_{m,n} \equiv \hat{\epsilon}_{m} -\hat{\epsilon}_{n}$ 
and excitation energies, 
\begin{eqnarray}
(\delta\hat{\epsilon}_{1,-2},\epsilon_{\rm exc}^{1\leftarrow -2})|^{N_{\rm f}=0}
&\stackrel{u=0}{\approx}& 
(105.95, 76.92),
\label{Ttwo_zero_uzero}
\\
(\delta\hat{\epsilon}_{2,1},\epsilon_{\rm exc}^{2\leftarrow 1})|^{N_{\rm f}=2}
&\stackrel{u=0}{\approx}& 
(105.08, 75.91),
\label{Ttwo_two_uzero}
\\
(\delta\hat{\epsilon}_{1,-2},\epsilon_{\rm exc}^{1\leftarrow -2})|^{N_{\rm f}=0}
&\stackrel{u=20}{\approx}& 
(93.31, 64.51),
\\
(\delta\hat{\epsilon}_{2,1},\epsilon_{\rm exc}^{2\leftarrow 1})|^{N_{\rm f}=2}
&\stackrel{u=-20}{\approx}& 
(93.43,  64.51)|^{K'},
\\
(\delta\hat{\epsilon}_{1,-2},\epsilon_{\rm exc}^{1\leftarrow -2})|^{N_{\rm f}=0}
&\stackrel{u=-20}{\approx}& 
(119.54, 90.28)|^{K'},
\\
(\delta\hat{\epsilon}_{2,1},\epsilon_{\rm exc}^{2\leftarrow 1})|^{N_{\rm f}=2}
&\stackrel{u=20}{\approx}& 
(117.75,  88.94),
\end{eqnarray}
in units of meV.
These level gaps and CR spectra at $N_{\rm f}$= (0,2) are related 
via $e$-$h$ conjugation, 
\begin{equation}
(\delta\hat{\epsilon}_{1,-2},\epsilon_{\rm exc}^{1\leftarrow -2})|^{N_{\rm f}=0}
= (\delta \hat{\epsilon}_{2,1},\,  \epsilon_{\rm exc}^{2\leftarrow1})|_{-\mu,-d,-r}^{N_{\rm f}=2}.
\end{equation}
In view of this, Eqs.~(\ref{Ttwo_zero_uzero}) and~(\ref{Ttwo_two_uzero}) imply 
that the effect of $e$-$h$ breaking $(d,r)$ is on the order of 1\%  in these level and CR spectra.
This is further confirmed by the approximate equality between the $u=\pm 20$\,meV expressions. 
It is somewhat surprising that the level gaps and CR spectra are only slightly affected 
by $e$-$h$ breaking $(d,r)$ 
while the level spectra $\{\hat{\epsilon}_{n} \}$ themselves are considerably modified 
(by a few \% generally, and much more for $\hat{\epsilon}_{1}$),
as noted in a paragraph below Eq.~(\ref{Spec_pzm_full}).
Accordingly, in experiment, the CR signals of $T_{2}$ will look nearly symmetric in $\nu$ 
at even-integer fillings.  
A shift or splitting of signals with $\nu$ around odd-integer filling $\nu \sim \pm 1$ and $\nu \sim \pm 3$ 
will reveal the presence or absence of orbital mixing.

It is seen from Eq.~(\ref{Ttwo_zero_uzero}) that Coulombic attraction accounts for roughly 30\% of 
the relevant level gap here for $T_{2}$. It diminishes rapidly for higher $T_{n}$;
it amounts to about 12\% for $T_{3}$  and 8\% for $T_{4}$.


\begin{figure}[btp]
\begin{center}
\includegraphics[scale=.72]{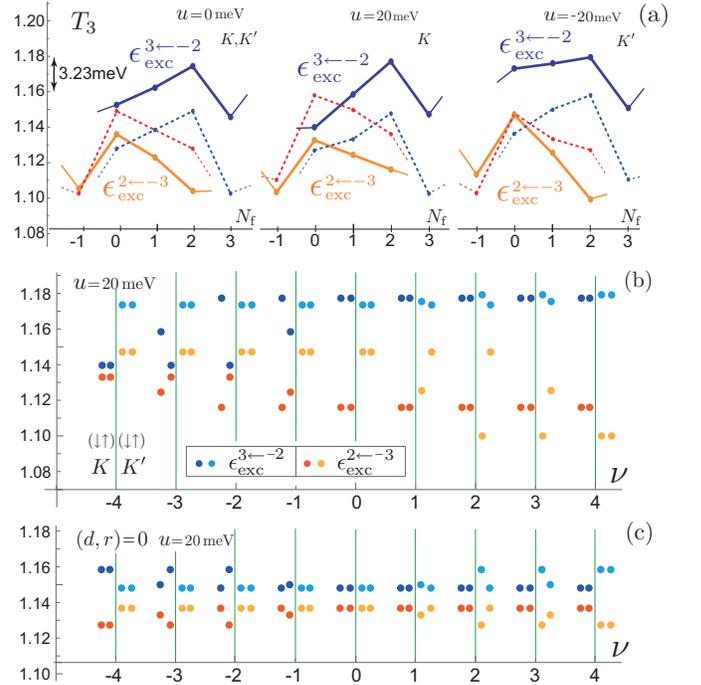}
\end{center}
\vskip-.7cm
\caption{  $T_{3}$ spectra  $(\epsilon_{\rm exc}^{3\leftarrow -2}, 
\epsilon_{\rm exc}^{2\leftarrow -3})$, 
normalized to the zero-$th$ level gap 
$(\epsilon_{3}^{(0)} -\epsilon_{-2}^{(0)}) \approx 1.18\,  \omega_{c}$, 
for $\tilde{V}_{c}/\omega_{c}=0.4$ at $B=$20\,T. 
(a)~Evolution of the $T_{3}$ spectra 
with filling $N_{\rm f}$ at bias $u=(0, \pm 20)$ meV.
Thin dotted lines guide the spectra of the $e$-$h$ symmetric case $(d, r)=0$.
(b)~Evolution of the full $T_{3}$ spectra in eight \{spin,valley\} channels with filling of the LLL 
according to Eq.~(\ref{fillingLLL_valley}).
(c)~For  $(d,r)=0$ the $T_{3}$ spectra are naturally symmetric in $\nu$.
}
\end{figure}


\subsection{Interband resonance $T_{3} \sim T_{5}$ }


\begin{figure}[tbp]
\begin{center}
\includegraphics[scale=.88]{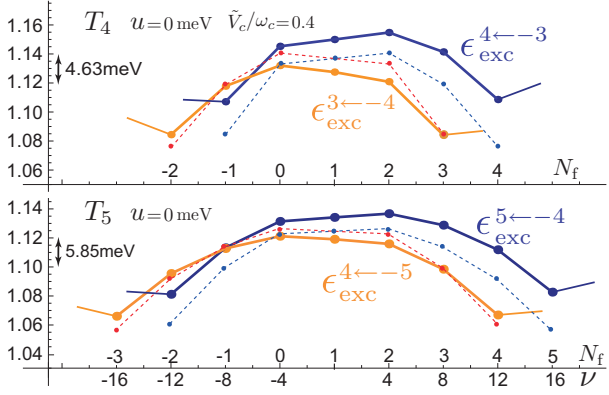}
\end{center}
\vskip-.7cm
\caption{
Resonance spectra $T_{4}$ and $T_{5}$, normalized to 
$(\epsilon_{4}^{(0)}\! -\epsilon_{-3}^{(0)}, \epsilon_{5}^{(0)}\! -\epsilon_{-4}^{(0)}\}
\approx (1.69, 2.13)\, \omega_{c}$,
for $\tilde{V}_{c}/\omega_{c}=0.4$ at $B$=20\,T.
$\nu$ refers to the total filling factor, $4(N_{\rm f}-1)$, to be 
realized when $N_{\rm f}$ is common to all spins and valleys.  
}
\end{figure}


Let us next consider other interband CR, $T_{3} \sim T_{5}$.
Figure 5(a) shows how the competing $T_{3}$ spectra 
$(\epsilon_{\rm exc}^{3\leftarrow -2}, \epsilon_{\rm exc}^{2\leftarrow -3})$
evolve with filling $N_{\rm f}$ of the relevant valley at bias $u=(0, \pm 20)$ meV.
Thin dotted lines guide the spectra of the $(d,r)=0$ case; 
they are symmetric about $N_{\rm f}=1$ at zero bias $u=0$.
As $e$-$h$ breaking $(d,r)$ is turned on, 
$\epsilon_{\rm exc}^{3\leftarrow -2}$ and $\epsilon_{\rm exc}^{2\leftarrow -3}$ 
deviate upward and downward, respectively,  by roughly 2~\% from the symmetric spectra, 
making the spectra split more on the $N_{\rm f}>1$ side and less on the other.

Figure~5(b) illustrates how the $2\times 2_{\rm spin}\times 2_{\rm valley}=8$ channels 
of $T_{3}$ spectra evolve over the range $-4 \le \nu \le 4$ at bias $u=20$ meV,
with the valley-dominant filling sequence of Eq.~(\ref{fillingLLL_valley}) assumed.
Splittings among the $(\epsilon_{\rm exc}^{3\leftarrow -2}, \epsilon_{\rm exc}^{2\leftarrow -3})$ spectra, 
unlike $T_{2}$, develop with $\nu$ noticeably, 
and $e$-$h$ breaking $(d,r)$ makes the full spectra visibly asymmetric in $\nu$, 
in contrast to the $e$-$h$ symmetric $[(d,r)=0]$ spectra, depicted in Fig.~5(c).

Figure 6 shows the CR spectra of 
$T_{4}=(\epsilon_{\rm exc}^{4\leftarrow -3}, \epsilon_{\rm exc}^{3\leftarrow -4})$ and 
$T_{5}=(\epsilon_{\rm exc}^{5\leftarrow -4}, \epsilon_{\rm exc}^{4\leftarrow -5})$ per spin and valley.
Included in the figure are the values of total filling factor $\nu$ to be realized 
when $N_{\rm f}$ (except $N_{\rm f}=1$) is common to all valleys and spins, 
i.e., $\nu = 4(N_{\rm f} -1)$.
These higher-energy resonances are practically insensitive to the detailed structure 
of the LLL and to a bias of $u\sim 20$ meV.
Again the CR spectra are less affected by $(d,r)\not=0$ than level shifts $\Delta \epsilon_{n}$, 
but are made visibly asymmetric in $\nu$.
The competing spectra overlap around $N_{\rm f}\sim -1$ or $\nu \sim -8$ 
and split more and more as $\nu \rightarrow 6$ for $T_{4}$ and $\nu \rightarrow 12$ for $T_{5}$.

Presumably, when $\nu$ changes over a wide range as in Fig.~6, 
screening of the Coulomb potential $v_{\bf p}$ will become important, 
as noted~\cite{KS_CRnewGR,SL} for monolayer graphene. 
Via screening $v_{\bf p}$ will get weaker with increasing $|\nu|$, 
making the spectra decrease faster for larger $|\nu|$.

\section{Summary and discussion}

Characteristic to bilayer graphene in a magnetic field is an octet of PZM levels 
nearly degenerate in orbitals  $n=\{0,1\}$ as well as in spins and valleys.
In this paper we have studied some basic characteristics of such PZM levels 
and shown that they generally undergo mixing in orbitals $\{0,1\}$ 
as they are gradually filled with electrons.
We have examined possible consequences  of orbital mixing 
and how they are detected by an observation of some leading competing channels 
of interband CR over a finite range of filling factor $\nu$.

It will be illuminating to summarize here why and how orbital mixing arises.
Let us first suppose an $e$-$h$ symmetric setting 
with $(d,r) \rightarrow 0$ and $u\rightarrow 0$. 
Coulomb interactions then lead to the orbital Lamb shift.
The resulting shift $\hat{\epsilon}_{0}- \hat{\epsilon}_{1}$ is odd 
under $e$-$h$ conjugation and changes sign 
as one goes from the empty to filled PZM sector, 
$(\hat{\epsilon}_{0}- \hat{\epsilon}_{1})|^{N_{\rm f}=0} 
=-(\hat{\epsilon}_{0}- \hat{\epsilon}_{1})|^{N_{\rm f}=2} 
= \epsilon_{\rm Ls}={1\over{8}}C\tilde{V}_{c}$; 
this is because the zero-modes  are $e$-$h$ selfconjugate 
[so that $\hat{\epsilon}_{n}|^{N_{\rm f}=0} 
= - \hat{\epsilon}_{n}|_{-X}^{N_{\rm f}=2}$ for $n= (0,1)$]. 
It is this level inversion that drives orbital mixing with filling of the PZM sector, 
as discussed in Sec.~V.
When valley and $e$-$h$ breaking $(u, d,r)$ is turned on, 
the level shift takes a modified form    
$(\hat{\epsilon}_{0}- \hat{\epsilon}_{1})|^{N_{\rm f}} = (\pm1-\xi)\epsilon_{\rm Ls}$
at $N_{\rm f}=(0, 2)$, with $\xi =O(\mu,d,r)$, and begins to change with bias $u$ 
slightly and almost linearly.  
The bias $u$ thereby acquires a critical value $u^{\rm cr}$,
beyond which orbital mixing disappears. 
In this way, orbital mixing is driven by the orbital Lamb shift,
and generally takes place in either valley or both (per spin) as bias $u$ is varied.

In our analysis, special attention has been paid to $e$-$h$ conjugation 
and valley-interchange operations, 
which govern, as in Eqs.~(\ref{En-ehconj}) and (\ref{Eexc-ehconj}), 
the level and CR spectra in bilayer graphene.  
In experiment, Coulombic corrections will be seen as variations 
of the interband CR spectra with filling $\nu$, 
$e$-$h$ breaking as an asymmetry of the spectra about $\nu=0$, 
and valley breaking as variations or splitting of the spectra with bias $u$.
Orbital mixing will be detected by an opening of some additional channels of CR.
The $T_{2}$ channels of interband CR, in particular, will serve 
as a direct and most sensitive probe 
to explore the novel characteristics of the LLL in bilayer graphene.


\appendix

\section{Counterterms $\delta \epsilon_{n}$}

In Sec.~IV the bare level spectra are written as $\epsilon_{n}=   \epsilon_{n}^{\rm ren} + \delta \epsilon_{n}$.
In this appendix we outline how to calculate the counterterms $\delta \epsilon_{n}$ numerically.
For given $\epsilon_{n}^{\rm ren}$,  one can write the associated counterterms as
\begin{equation}
\delta \epsilon_{n} = \delta_{\rm ct} \epsilon_{n}^{\rm ren} 
\equiv \lambda_{n}\,  \delta v/v^{\rm ren},  
\end{equation}
with the differential operator 
\begin{equation}
\delta_{\rm ct} = \delta v{\partial\over{\partial v^{\rm ren}}} 
+ \delta \gamma_{1} {\partial\over{\partial \gamma_{1}^{\rm ren}}} 
+\delta \Delta {\partial\over{\partial\Delta^{\rm ren}}}  
\end{equation}
acting on $\epsilon_{n}^{\rm ren}$ and with $(\delta v, \delta \gamma_{1}, \delta \Delta)$ defined in  Eq.~(\ref{deltagamma}). This formula allows one to  
evaluate $\lambda_{n}$ analytically. 
Alternatively, one can let $\delta_{\rm ct}$ act on the reduced matrix 
$\hat{\cal H}_{N}$ in Eq.~(\ref{reducedH}), and write 
\begin{equation}
\delta  \epsilon_{n} = ({\bf b}_{n})^{\dag}\! \cdot 
\delta_{\rm ct} \hat{\cal H}_{N}\! 
\cdot  {\bf b}_{n}
= \lambda_{n}\, \delta v/v^{\rm ren}.
\label{matrixct}
\end{equation}
Actually,  $\Lambda_{N}\equiv \delta_{\rm ct}\hat{\cal H}_{N}/(\delta v/v^{\rm ren})$ 
is given by $\hat{\cal H}_{N}$ with substitution 
$(\mu,r)\rightarrow 0$ first and subsequently $g \rightarrow (g + r\, d)\, h(r)$ and
 $d \rightarrow 2(d + r\, g) h(r)$.
In this way one can directly calculate $\lambda_{n} = {\bf b}_{n}^{\dag}\! \cdot\! \Lambda_{N}\! 
\cdot\! {\bf b}_{n}$ from eigenmodes ${\bf b}_{n}$.

\section{Resonance Spectra $T_{2}$}

In this Appendix, we outline the derivation of the CR spectra 
$T_{2}=\{2\leftarrow {\rm PZM},\  {\rm PZM} \leftarrow -2\}$
examined in Sec.~VI.  
Let us first write the renormalized level spectra as
$\hat{\epsilon}_{n}= \hat{\epsilon}_{n}|^{N_{\rm f}=0} 
+ \delta\epsilon_{n}^{\rm SE}$.
As the PZM sector is gradually filled over the range $0\le N_{\rm f}=N_{1}+ N_{0} \le 2$, 
each level $n$ acquires an additional self-energy correction
\begin{equation}
\delta \epsilon_{n}^{\rm SE}
=-  \sum_{\bf p} v_{\bf p} \gamma_{\bf p}^2 \, 
\{ N_{1} |G^{n1}_{\bf p}|^2 + N_{0} |G^{n0}_{\bf p}|^2\}.
\end{equation}
Direct calculation of $\delta \epsilon_{0}^{\rm SE}$ and $\delta \epsilon_{1}^{\rm SE}$ 
leads to $(\hat{\epsilon}_{0_{\theta}},  \hat{\epsilon}_{1_{\theta}})$ in Eq.~(\ref{EPZM-theta}).
Adding Coulombic attraction terms
\begin{equation}
{\cal A}_{nj}
=\sum_{\bf p} v_{\bf p} \gamma_{\bf p}^2 \, G^{nn}_{\bf p} G^{jj}_{\bf -p} 
\end{equation}
then yields the CR spectra
$\epsilon_{\rm exc}^{n\leftarrow j} 
= \hat{\epsilon}_{n}- \hat{\epsilon}_{j} -  (\nu_{j}-  \nu_{n} )\, {\cal A}_{n,j}$.

When orbital mixing is present, i.e., for $u<u^{\rm cr}$, 
interband CR, $T_{2}$, hosts four active channels (per spin and valley) 
over the range $N_{\rm f}^{-} < N_{\rm f} < N_{\rm f}^{+}$.
Associated with the $n =1_{\theta}$ level are the excitation spectra,
\begin{eqnarray}
\epsilon_{\rm exc}^{2 \leftarrow 1_{\theta}}
&=& \hat{\epsilon}_{2} -\hat{\epsilon}_{1_{\theta}} - N_{1}\, {\cal A}_{21},
\label{Eapp1}
\\
\epsilon_{\rm exc}^{1_{\theta} \leftarrow -2}
&=& \hat{\epsilon}_{1_{\theta}}  - \hat{\epsilon}_{-2}\! - (1-N_{1})\, {\cal A}_{1,-2},
\end{eqnarray}
and those associated with the $n= 0_{\theta}$ level are
\begin{eqnarray}
\epsilon_{\rm exc}^{2 \leftarrow 0_{\theta}}
&=&  \hat{\epsilon}_{2} - \hat{\epsilon}_{0_{\theta}}  - N_{0}\, {\cal A}_{20},
\\
\epsilon_{\rm exc}^{0_{\theta} \leftarrow -2}
&=& \hat{\epsilon}_{0_{\theta}} - \hat{\epsilon}_{-2} -(1- N_{0})\, {\cal A}_{0,-2}.
\label{Eapp2}
\end{eqnarray}
In actual calculations one can simplify, 
for $n\not =  (0,1)$, 
$|G^{n1}|^2 \rightarrow c_{\theta}^{2} |g^{n1}|^2 + s_{\theta}^{2} |g^{n0}|^2$, 
$G^{nn} G^{11} \rightarrow g^{nn}  (c_{\theta}^{2}\,  g^{11} + s_{\theta}^{2}\, g^{00})$,
etc., under symmetric integration $\sum_{\bf p}$. 
Evaluating Eqs.~(\ref{Eapp1}) $\sim$ (\ref{Eapp2}) numerically leads 
to the excitation spectra depicted in Fig.~4.




\begin{thebibliography}{99}
\bibitem{fn}  In ref.~\cite{KS_CR_eh} we adopted a slightly different set of band parameters.
 All our numerical analysis in the present paper 
 has also been made with it and essentially the same conclusion has been reached.
\end{thebibliography}
\end{document}